\documentclass[a4paper,fleqn,usenatbib]{mnras}

\usepackage{newtxtext,newtxmath}
\usepackage[T1]{fontenc}
\usepackage{ae,aecompl}

\usepackage{graphicx}	% Including figure files
\usepackage{amsmath}	% Advanced maths commands
\usepackage{amssymb}	% Extra maths symbols

\title[TTVs and refined system parameters of XO-6b]{Periodic transit timing variations and refined system parameters of the exoplanet XO-6b}

\author[Z. Garai et al.]{
Zolt\'an Garai,$^{1,2,3}$\thanks{E-mail: zgarai@ta3.sk}
Theodor Pribulla,$^{3,2,1}$
Richard Kom\v{z}\'{i}k,$^{1}$
Emil Kundra,$^{1}$
\newauthor
\v{L}ubom\'{i}r Hamb\'{a}lek,$^{1}$
and Gyula M. Szab\'{o}$^{2,3}$\\
$^{1}$Astronomical Institute, Slovak Academy of Sciences, 05960 Tatransk\'a Lomnica, Slovakia\\
$^{2}$MTA-ELTE Exoplanet Research Group, 9700 Szombathely, Szent Imre h. u. 112, Hungary\\
$^{3}$ELTE Gothard Astrophysical Observatory, 9700 Szombathely, Szent Imre h. u. 112, Hungary 
}

\date{Accepted XXX. Received YYY; in original form ZZZ}

\pubyear{2019}

\begin{document}
\label{firstpage}
\pagerange{\pageref{firstpage}--\pageref{lastpage}}
\maketitle

% Abstract of the paper
\begin{abstract}
Only a few exoplanets are known to orbit around fast rotating stars. One of them is XO-6b, which orbits an F5V-type star. Shortly after the discovery, we started multicolor photometric and radial-velocity follow-up observations of XO-6b, using the telescopes of Astronomical Institute of the Slovak Academy of Sciences. Our main scientific goals were to better characterize the planetary system and to search for transit timing variations. We refined several planetary and orbital parameters. Based on our measurements, the planet XO-6b seems to be about 10\% larger, which is, however, only about $2\sigma$ difference, but its orbit inclination angle, with respect to the plane of the sky, seems to be significantly smaller, than it was determined originally by the discoverers. In this case we found about $9.5\sigma$ difference. Moreover, we observed periodic transit timing variations of XO-6b with a semi-amplitude of about 14 min and with a period of about 450 days. There are two plausible explanations of such transit timing variations: (1) a third object in the system XO-6 causing light-time effect, or (2) resonant perturbations between the transiting planet XO-6b and another unknown low-mass planet in this system. From the O-C diagram we derived that the assumed third object in the system should have a stellar mass, therefore significant variations are expected in the radial-velocity measurements of XO-6. Since this is not the case, and since all attempts to fit radial velocities and O-C data simultaneously failed to provide a consistent solution, more realistic is the second explanation. 
\end{abstract}

% Select between one and six entries from the list of approved keywords.
% Don't make up new ones.
\begin{keywords}
planets and satellites: individual: XO-6b -- techniques: photometric -- techniques: radial velocities -- methods: observational
\end{keywords}

%%%%%%%%%%%%%%%%%%%%%%%%%%%%%%%%%%%%%%%%%%%%%%%%%%

%%%%%%%%%%%%%%%%% BODY OF PAPER %%%%%%%%%%%%%%%%%%

\section{Introduction}

Based on the on-line exoplanet catalogs\footnote{See for example \url{https://exoplanetarchive.ipac.caltech.edu} or \url{http://exoplanet.eu}.}, more than 4120 confirmed exoplanets in 3063 planetary systems were discovered up to November 2019. The total census of transiting exoplanets is more than 2960 planets in 2223 planetary systems. Up to this date, the most successful transiting-exoplanet searcher is the \textit{Kepler} space telescope \citep{Borucki2, Borucki3, Borucki1}, launched in 2009 and decomissioned in 2018. The \textit{Kepler} and \textit{K2} missions \citep{Howell1} together discovered 2734 confirmed transiting exoplanets. On the other hand, Kepler parent stars are relatively faint in general and, therefore, difficult for ground-based radial velocity (RV) follow-up observations. This is the reason, why other methods were used for confirmation of \textit{Kepler} exoplanet candidates: for example, confirmation from induced variability \citep{Shporer1}, from multi-planet statistics \citep{Lissauer1}, or from imaging \citep{Law1}. This is also the reason, why the all-sky TESS mission \citep{Ricker1}, launched in 2018, and the planned CHEOPS mission \citep{Broeg1}, targeted bright stars. 

To get true masses of the exoplanets via RV measurements \citep{Mayor1} we need planets transiting in front of bright stars. Interesting, but not surprising is the fact that the vast majority of these systems in the \textit{Kepler} era was discovered not by the above mentioned \textit{Kepler} and \textit{K2} missions, but by ground-based transit surveys, like HATNet \citep{Bakos1}, SuperWASP \citep{Street1}, or XO \citep{McCullough1}. For our follow-up observations, we selected the transiting exoplanet XO-6b, discovered recently by \citet{Crouzet1} within the framework of the XO project, which started in 2005. In 2011, a new version of XO instrument was developed and installed at three observatories: (1) at Vermillion Cliffs Observatory in USA, (2) at Observatorio del Teide in Spain, and (3) at Observatori Astron\'{o}mic del Montsec, also in Spain. Each XO observatory consists from two 10 cm diameter and 200 mm focal length Canon telephoto lenses, equipped with an Apogee E6 $1024 \times 1024$ pixels CCD camera, mounted on a German-Equatorial Paramount ME mount. All six lenses and cameras operate in a network configuration, scanning the same segment of the sky. The XO project discovered six transiting exoplanets up to November 2019 \citep{Crouzet2}, the last one is XO-6b, which orbits a relatively bright ($V \approx 10.2$ mag), hot ($T_\mathrm{eff} \approx 6720$ K), and fast rotating ($v~\mathrm{sin}~i \approx 48$ km s$^{-1}$) parent star XO-6 (TYC 4357-995-1). Its equatorial coordinates are the following: R.A. 06:19:10.3604 and Dec. +73:49:39.602 (J2000.0). The mass of the star is $M_\mathrm{s} = 1.50(3)$~M$_{\odot}$ and its radius is $R_\mathrm{s} = 1.86(16)$~R$_{\odot}$. The exoplanet XO-6b is also very interesting, since it is a transiting hot Jupiter with the mass of $M_\mathrm{p} < 4.4$~M$_\mathrm{Jup}$ and the radius of $R_\mathrm{p} = 2.07(22)$~R$_\mathrm{Jup}$ on a prograde and misaligned orbit, with the $\lambda = -20.7(2.3)$ deg, which is the sky-projected angle between the stellar spin axis and the planetary orbit axis. The orbital period of the planet, $P_\mathrm{orb} = 3.765000(8)$ days, is longer than the rotation period of the parent star, $P_\mathrm{rot} < 2.12$ days \citep{Crouzet1}.

In 2017, shortly after the discovery, we started multicolor photometric and RV follow-up observations of the newly discovered transiting exoplanet XO-6b, with the main scientific goals to better characterize the planetary system and to search for transit timing variations (TTVs). Our motivations were the following: (1) As of November 2019, only discovery and initial follow-up observations published by \citet{Crouzet1} are available. We decided to carry out an independent follow-up campaign, with the time-base of at least 1 year. (2) The discoverers (C17) noted that an extensive RV campaign would be necessary to measure the mass of the planet with greater confidence. Only the $3\sigma$ upper limit on planet's mass was derived up to now, which is $4.4$~M$_\mathrm{Jup}$. The precision of RV measurements is negatively influenced by the large rotational velocity of the parent star. (3) XO-6b is a planet, which orbits and transits a fast rotating parent star. Only a few similar exoplanets are known. (4) C17 found no evidence for TTVs. On the other hand, if we take a look at the O-C diagram of XO-6b transits, presented in the Exoplanet Transit Database\footnote{See \url{http://var2.astro.cz/ETD/}.} (ETD), this diagram shows clear TTVs \citep{Poddany1}. The no detection of C17 can be due to the relatively short time span and not optimal time sampling of their observations, and/or higher uncertainties in their O-C values. 

TTVs were identified only in the case of a few hot Jupiters. The suspected outer companions are planets and brown dwarfs, and in one example, a multi-planet system is suspected in the outskirts. In some cases, the outer companions are also known to exhibit their signal in the RV measurements, as well, while distant companions of other exoplanets are only known by RV solutions. \citet{Nascimbeni1} observed a non-periodic TTV in WASP-3b transits with an unknown origin, while the likely explanations include a possible multi-planet outer system. \citet{Maciejewski1} give a joint solution of photometric TTVs and RV measurements of the WASP-12 system, leading to a solution with a $0.1$~M$_\mathrm{Jup}$ mass exoplanet. \citet{Neveu-VanMalle1} reports that RV measurements of WASP-47 is compatible with an outer companion of a similar mass than the transiting exoplanet, in the order of $1$~M$_\mathrm{Jup}$. In the case of WASP-41, the scatter in RV fitting reduces to 1/10th of its value when adding a second planet to the fit, which is an evidence for an outer planet in WASP-41, as well. \citet{Dawson1} give a solution to KOI-1747.01, concluding on an outer brown dwarf with a $24$~M$_\mathrm{Jup}$ mass. \citet{Knutson1} reported that the second companion in three exoplanets, including HAT-P-13, are suspected brown dwarfs. WASP-8 is a double system, derived from the RV acceleration, with no mass solution for the "c" planet. Similarly, the outer companion of WASP-34b is likely a brown dwarf. \citet{Hartman1} reported about an increased scatter of HAT-P-44 and HAT-P-46 that can be explained by a 2-planet fit and a possible 2-planet solution is also suggested to HAT-P-45, with less confidence based on the available data. Some known examples of TTVs has recently been debated. \citet{Seeliger1} show that suspected TTVs in the transit O-C of HAT-P-32b can in fact be explained assuming only one planet, after improving the linear fit. \citet{Wang1} show that observations of HAT-P-25b is also compatible to a single planet solution. These examples are, however, relatively rare to the number of known hot Jupiters, and because of this rarity, a bimodal formation of close planet systems has been suggested, see \citet{Sandford1} and references therein. As possible outcomes, one major hot Jupiter alone, or several lower-mass planets on either close-in and/or more distant orbits. This model is successful in explaining the structure of the distribution of transiting planets and is compatible with the current understanding of planet formation, as well. However, each system, where a hot Jupiter is detected with an outer, especially planet-mass companion, is a challenge that has to be explained as an exception. This is why the characterization of the
transiting planets and the TTV/RV measurements is a central question for all the known cases.

The paper is organized as follows. In Section \ref{obs} a brief introduction to instrumentation and data reduction is given. The follow-up photometry and spectroscopy is detailed in Section \ref{followup}. The analysis of TTVs is described in Section \ref{ttv}. Our findings are summarized in Section \ref{conc}. 

\section{Instrumentation, observations, and data reduction}
\label{obs}

The follow-up observations were performed at Astronomical Institute of the Slovak Academy of Sciences\footnote{See \url{https://www.ta3.sk/}.} (AI SAS). The institute operates several telescopes at different observing stations. Our multicolor photometric observations were performed mainly at the Star\'{a} Lesn\'{a} Observatory of AI SAS, in its G2 observing pavilion. It has the geographical coordinates of 20$\degr$17'25"E and 49$\degr$09'10"N, and it is located at 810 m a.s.l. Photometric observations were also obtained at the Skalnat\'{e} Pleso Observatory of AI SAS. It has the coordinates of 20$\degr$14'02"E and 49$\degr$11'22"N, and it is located at 1786 m a.s.l. The spectroscopic observations were obtained at the Skalnat\'{e} Pleso Observatory only. 

\subsection{Photometric observations and data reduction}
\label{phot}

Our photometric observations were obtained between March 2017 and March 2019. We started the monitoring of XO-6b transits at the G2 pavilion of Star\'{a} Lesn\'{a} Observatory, using the 0.6 m (f/12.5) Zeiss Equatorial Cassegrain reflecting telescope, equipped with an FLI ML-3041 $2048 \times 2048$ pixels CCD camera. This instrument was already used with success, for example, during the YETI campaigns on NGC 7243 \citep{Garai2}. At the Skalnat\'{e} Pleso Observatory (SP) we used the newly installed 1.3 m (f/8.36) Astelco Alt-azimuthal Nasmyth-Cassegrain reflecting telescope. At this observing site we used an MI G4-9000 $3048 \times 3048$ pixels CCD camera at the first observation (See Table \ref{photobslog}), later the photometric detector was changed to an FLI PL-16803 $4096 \times 4096$ pixels CCD camera. We alternated Johnson-Cousins \textit{B}, \textit{V}, \textit{R}, and \textit{I} filters during the observations \citep{Bessell1}, and since the target star is relatively bright ($V \approx 10.2$ mag), we applied quite short integrations of 20 sec, 20 sec, 10 sec, and 10 sec in \textit{B}, \textit{V}, \textit{R}, and \textit{I} passbands, respectively. We observed 11 individual transit events in total, but, unfortunately, we had to discard 3 photometric measurements during the analysis, due to the low data quality. The log of the remaining 8 multicolor photometric observations used in our analysis is given in Table \ref{photobslog}.        

\begin{table}
\centering
\caption{Log of multicolor photometric observations used in our analysis. Table shows number of scientific frames per passband, obtained during the given observing night at given observatory: G2 -- Star\'{a} Lesn\'{a} Observatory, G2 pavilion; SP -- Skalnat\'{e} Pleso Observatory.}
\label{photobslog}
\begin{tabular}{cccccc}
\hline
\hline
Evening date &    \textit{B}    &  \textit{V}   &  \textit{R}   &   \textit{I}    & Observatory\\
\hline
2017-03-14   &    269  &  253 &  238 &   256  &  G2\\
2017-05-17   &    189  &  191 &  193 &   188  &  G2\\
2017-06-01   &    77   &  77  &  86  &   81   &  SP\\
2017-07-20   &    224  &  213 &  209 &   221  &  G2\\
2018-04-02   &    209  &  201 &  198 &   209  &  G2\\
2018-08-08   &    276  &  232 &  267 &   275  &  G2\\
2018-10-11   &    120  &  122 &  122 &   123  &  G2\\
2019-03-22   &    167  &  124 &  161 &   0    &  SP\\
\hline
Frames total &    1531 & 1413 & 1474 &   1353 & 5771\\
\hline       
\hline
\end{tabular}
\end{table}   

The obtained photometric observations were reduced using the software package {\tt{IRAF}}\footnote{See \url{https://github.com/iraf-community/iraf}.}. The reduction included dark and flat corrections, astrometric calibrations and differential photometry. We applied differential photometry in 2 modifications. As a first modification we applied differential photometry using a comparison star and a check star. For this method we selected the closest stars of similar brightness. We used the star 2MASS J06190411+7349116 as a comparison star, and the star 2MASS J06182685+7350022 as a check star. As a second modification we applied differential photometry using an artificial standard star as described by \citet{Broeg2}. In this case, all stars in the field were used to create an artificial standard star, but the more photometrically stable stars were assigned higher weights. This artificial standard star was used to calculate the differential magnitudes of all objects, including XO-6. Finally, we compared the results from the first and the second modifications of differential photometry and selected the light curves with smaller scatter. Subsequently, we converted all time-stamps from Heliocentric Julian Date (HJD) to Barycentric Julian Date in Barycentric Dynamical Time (BJD$_\mathrm{TDB}$), using the on-line applet {\tt{HJD2BJD}}\footnote{See \url{http://astroutils.astronomy.ohio-state.edu/time/hjd2bjd.html}.} \citep{Eastman2}. After the linear trend, due to the second-order extinction, was removed from the photometric data, we cleaned light curves from outlier data points. We used a $3\sigma$ clipping (where $\sigma$ is the standard deviation of the light curve), according to effective cleaning, as well as in order to avoid cleaning in-transit points. Resulting individual light curves were used during the next analysis (See Sect. \ref{followup}).

\subsection{Spectroscopic observations and data reduction}      
\label{spec}

Our spectroscopic observations were obtained between September 2018 and February 2019 at the Skalnat\'{e} Pleso Observatory, using the 1.3 m telescope (See Sect. \ref{phot}), equipped with a fiber-fed echelle spectrograph of MUSICOS design \citep{Baudrand1}. Its fiber injection and guiding unit (FIGU) is mounted in the second Nasmyth focus of the telescope. The FIGU is connected to the calibration unit (ThAr hollow
cathode lamp, tungsten lamp, blue LED) in the control room and to the echelle spectrograph itself in the room below the dome, where the temperature is stable. The spectra were recorded by an Andor iKon-936 DZH $2048 \times 2048$ pixels CCD camera. The spectral range of the instrument is 4250 -- 7375 \AA~ in 56 echelle orders. The maximum resolution of the spectrograph reaches $R \approx 38~000$ around 6000 \AA. The exposure time was 900 sec in all cases. Usually three raw spectra were obtained during an observing night and, subsequently, combined via {\tt{IRAF}} task {\tt{combine}} to increase the signal-to-noise ratio (SNR). Three 900-sec exposures correspond to about 0.27\% of the orbital period, so the orbital-motion smearing of spectral lines is negligible. 

The data were reduced using {\tt{IRAF}} package tasks, {\tt{Linux}} shell scripts, and {\tt{FORTRAN}} programs similarly, as it was described in \citet{Pribulla1} and in \citet{Garai1}. In the first step, master dark frames were produced. In the second step, the photometric calibration of the frames was done using dark and flat-field frames. Bad pixels were cleaned using a bad pixel mask, and cosmic hits were removed using the program of \citet{Pych1}. Photometrically calibrated frames were combined to increase the SNR. Order positions were defined by fitting Chebyshev polynomials to tungsten lamp and blue LED spectrum. In the following step, scattered light was modeled and subtracted. Aperture spectra were then extracted for both object and ThAr frames, and then the resulting 2D spectra were dispersion-solved. Two-dimensional spectra were finally combined to 1D spectra rebinned to 4250 -- 7375 \AA~ wavelength range with a 0.05 \AA~ step, i.e. about 2 -- 4 times the spectral resolution. Spectra were analyzed using the broadening function (BF) technique, developed by \citet{Rucinski1}, to get RVs. Because the formal RV uncertainties found using the BF approach are hard to quantify and depend on BF smoothing, they were determined from SNR as follows. First we determined RV uncertainties as 1/SNR \citep{Hatzes1}. Subsequently, we fitted the RV observations with initial uncertainties using the code {\tt{RMF}}, described in Sect. \ref{trandrvanalysis}, and from the best fit we obtained reduced $\chi^{2}$ ($\chi^{2}_\mathrm{red}$). In the next step, we rescaled all uncertainties to get $\chi^{2}_\mathrm{red} = 1$. SNR of the spectra can be obtained from the $1\sigma$ uncertainties as $C/\sigma$, where the scaling constant $C$ was found to be 21.5664 km~s$^{-1}$. Unfortunately, we had to discard several spectroscopic measurements due to problems with stability of the spectrograph. An overview of the remaining 7 RV values with uncertainties is shown in Table \ref{spectrobslog}. In these 7 cases, systematic errors caused by the spectrograph instability were on the level from 0.02 up to 0.1 km~s$^{-1}$.

\begin{table}
\centering
\caption{Log of spectroscopic observations, obtained at the Skalnat\'{e} Pleso Observatory. Table shows RV values and their $\pm 1\sigma$ uncertainties, sorted by the evening date.}
\label{spectrobslog}
\begin{tabular}{ccc}
\hline
\hline
Evening date &    RV values [km~s$^{-1}$]  &  $\pm 1\sigma$ [km~s$^{-1}$]\\
\hline
2018-09-12   &    $-4.5$             &  $0.5$\\
2018-09-19   &    $-4.5$             &  $0.6$\\
2018-09-20   &    $-4.7$             &  $0.6$\\
2019-02-15   &    $-3.9$             &  $0.8$\\
2019-02-16   &    $-3.0$             &  $0.7$\\
2019-02-17   &    $-3.5$             &  $0.6$\\
2019-02-18   &    $-3.3$             &  $0.6$\\
\hline
\hline
\end{tabular}
\end{table}   

\section{Analysis of follow-up observations of the XO-6 system}
\label{followup}

Following our main scientific goals, we analyzed the multicolor photometric data and RV observations in 2 steps. As first, in Sect. \ref{indtransitanalysis}, we analyzed the transits per passband. During the next step, in Sect. \ref{trandrvanalysis}, we simultaneously analyzed all transits and RV observations. Here, we describe our analysis procedure. Our results will be presented and discussed, as well.

\subsection{Analysis of transits per passband} 
\label{indtransitanalysis}

\begin{table*}
\centering
\caption{An overview of {\tt{JKTEBOP}} best-fit parameters. Lines marked with a $*$ are parameter values presented by \citet{Crouzet1}, which were listed only for easier comparison with the corresponding values from this work. The weighted averages were calculated with weights of $1/\sigma^2$, where $\sigma$ is the uncertainty in each passband.}
\label{paramperbands}
\begin{tabular}{cccccc}
\hline
\hline
Parameters 				&	\textit{B} 		&	\textit{V}		&	\textit{R}		&	\textit{I}		& Weighted average\\
\hline
$(R_\mathrm{p} + R_\mathrm{s})/a$ 	&	$0.155(11)$		&	$0.148(8)$ 		&	$0.146(7)$ 		&	$0.149(9)$		& $0.148(4)$\\
$^*a/R_\mathrm{s}$                      &       $9.2(4)$                &       $9.0(3)$                &       $9.0(4)$                &       $9.4(3)$                & $9.20(19)$\\   
$R_\mathrm{p}/R_\mathrm{s}$ 		&	$0.126(4)$		&	$0.118(2)$ 		&	$0.124(2)$ 		&	$0.123(2)$		& $0.1220(11)$\\
$^*R_\mathrm{p}/R_\mathrm{s}$		&	$0.117(2)$ 		&	$0.1151(17)$ 		&	$0.1153(19)$ 		&	$0.1114(17)$		& $0.1144(13)$\\
$i$ [deg]				&	$83.7(7)$		&	$83.9(5)$ 		&	$84.4(4)$ 		&	$84.0(6)$		& $84.0(2)$\\
$^*i$ [deg]				&	$85.9(5)$ 		&	$86.0(3)$		&	$85.9(4)$ 		&	$85.9(3)$		& $85.9(2)$\\
$P_\mathrm{orb}$ [days]			&	$3.765038(11)$		&	$3.765049(8)$		&	$3.764980(6)$		&	$3.764971(11)$		& $3.765004(4)$\\ 
$T_\mathrm{c} - 2~456~652$ [BJD]	&	$0.691(4)$ 		&	$0.689(4)$		&	$0.719(3)$ 		&	$0.720(4)$		& $0.7070(18)$\\
$L_\mathrm{sf}$ [mag]			&	$-0.0010(3)$		&	$-0.0012(3)$		&	$-0.0010(2)$		&	$-0.0004(3)$		& $-0.00092(13)$\\
\hline
\hline
\end{tabular}
\end{table*}   

As a first step we fitted the observed transits simultaneously in each passband, using the {\tt{JKTEBOP}}\footnote{See \url{https://www.astro.keele.ac.uk/~jkt/codes/jktebop.html}.} code \citep{Southworth1}, which is based on the former {\tt{EBOP}} code \citep{Popper1}. The {\tt{JKTEBOP}} code is used to fit a model to the light curves of detached eclipsing binary stars in order to derive the radii of the stars as well as various other quantities. It is very stable and has a lot of additional features, including extensive Monte Carlo or bootstrapping uncertainty analysis algorithms. It is also widely used for transiting extrasolar planetary systems. During the fitting procedure 6 free parameters were adjusted, including the sum of fractional radii $(R_\mathrm{p} + R_\mathrm{s})/a$, the planet-to-star radius ratio $R_\mathrm{p}/R_\mathrm{s}$, the orbit inclination angle of the planet $i$ with respect to the plane of the sky, the orbital period of the planet $P_\mathrm{orb}$, the mid-transit time $T_\mathrm{c}$, and the light-scale factor $L_\mathrm{sf}$. The starting values of the parameters were taken from \citet{Crouzet1}. Only the limb-darkening (LD) coefficients were fixed during the fitting procedure. We used the quadratic LD law. The corresponding coefficients were linearly interpolated based on the stellar parameters of $T_\mathrm{eff} = 6720$ K, log $g = 4.04$ (cgs), and Fe/H $= -0.07$ \citep{Crouzet1} for \textit{B}, \textit{V}, \textit{R}, or \textit{I} passband, using the on-line applet {\tt{EXOFAST -- Quadratic Limb Darkening}}\footnote{See \url{http://astroutils.astronomy.ohio-state.edu/exofast/limbdark.shtml}.}, which is based on the IDL-routine {\tt{QUADLD}} \citep{Eastman1}. This software interpolates the \citet{Claret1} quadratic limb-darkening tables. To estimate the uncertainties in the fitted parameters we used the {\tt{JKTEBOP}}-task No. 8, which executes Monte-Carlo simulations. This finds the best fit and then uses Monte-Carlo simulations to estimate the uncertainties in the parameters. The best fitting model is re-evaluated at the phases of the actual observations. Gaussian simulated observational noise is added, and the result is refitted. This process is repeated, and the range in parameter values found gives the uncertainty in that parameter. To obtain final results we produced 10~000 random data realizations.

\begin{figure*}
\centering
\centerline{
\includegraphics[width=88mm]{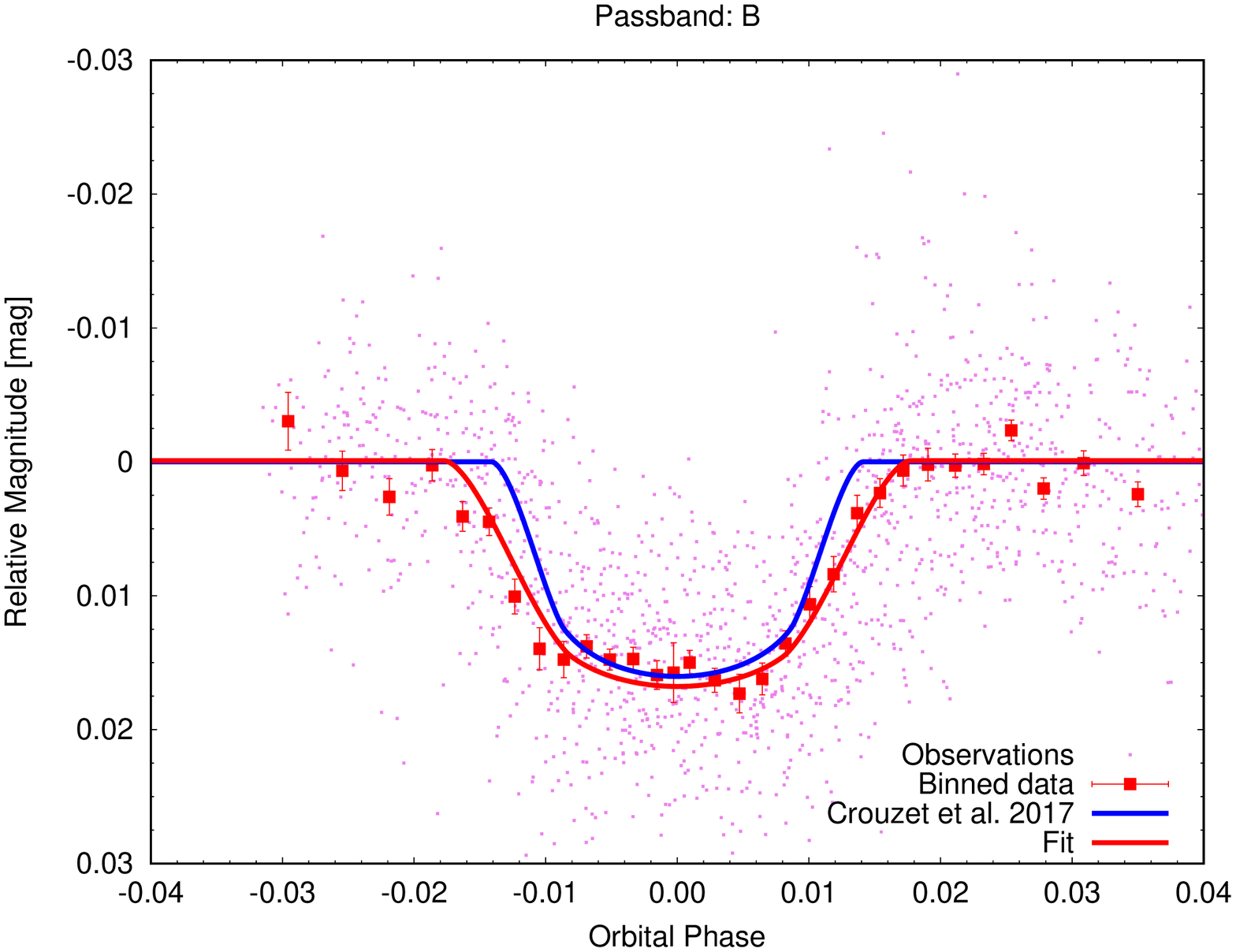}
\includegraphics[width=88mm]{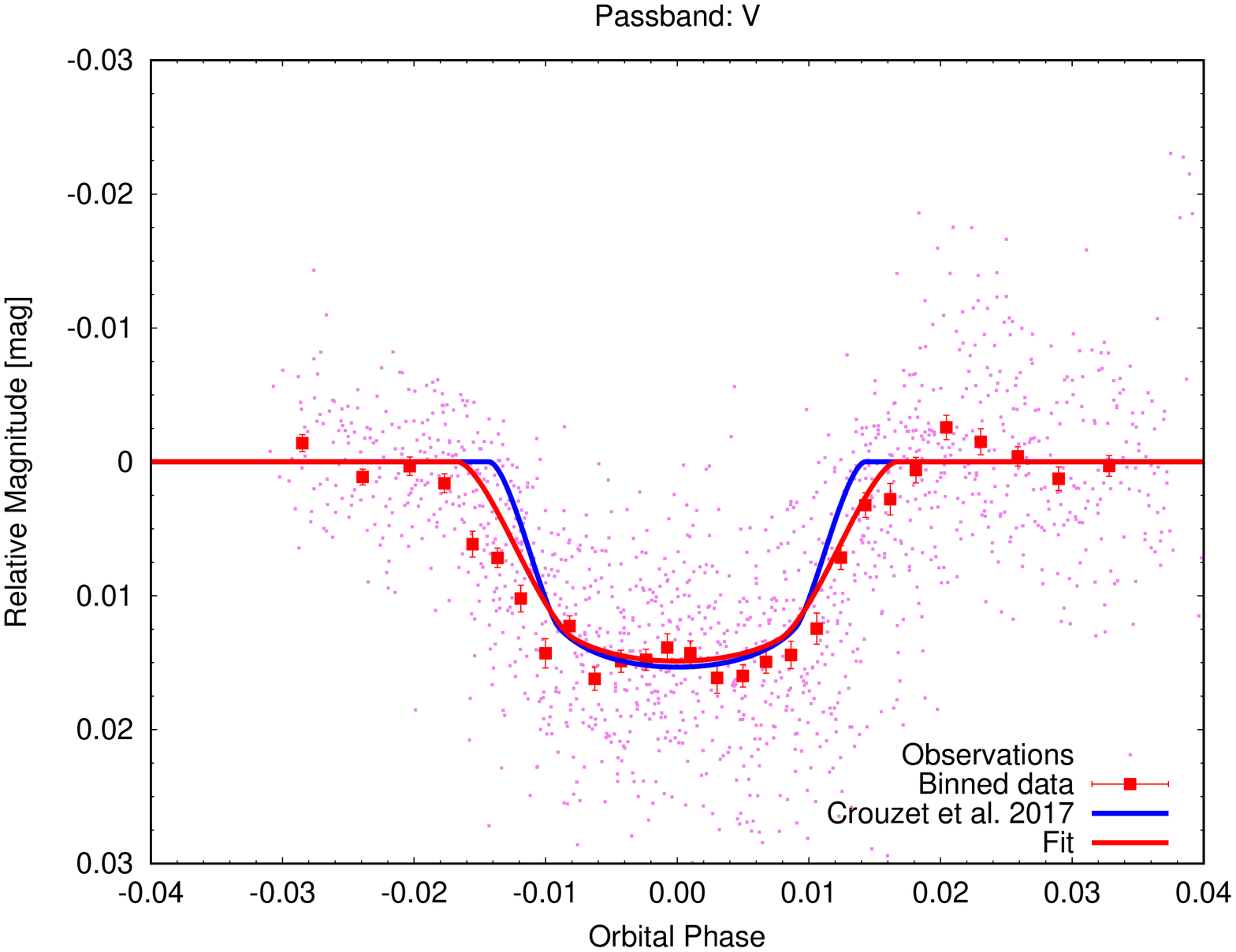}}
\centerline{
\includegraphics[width=88mm]{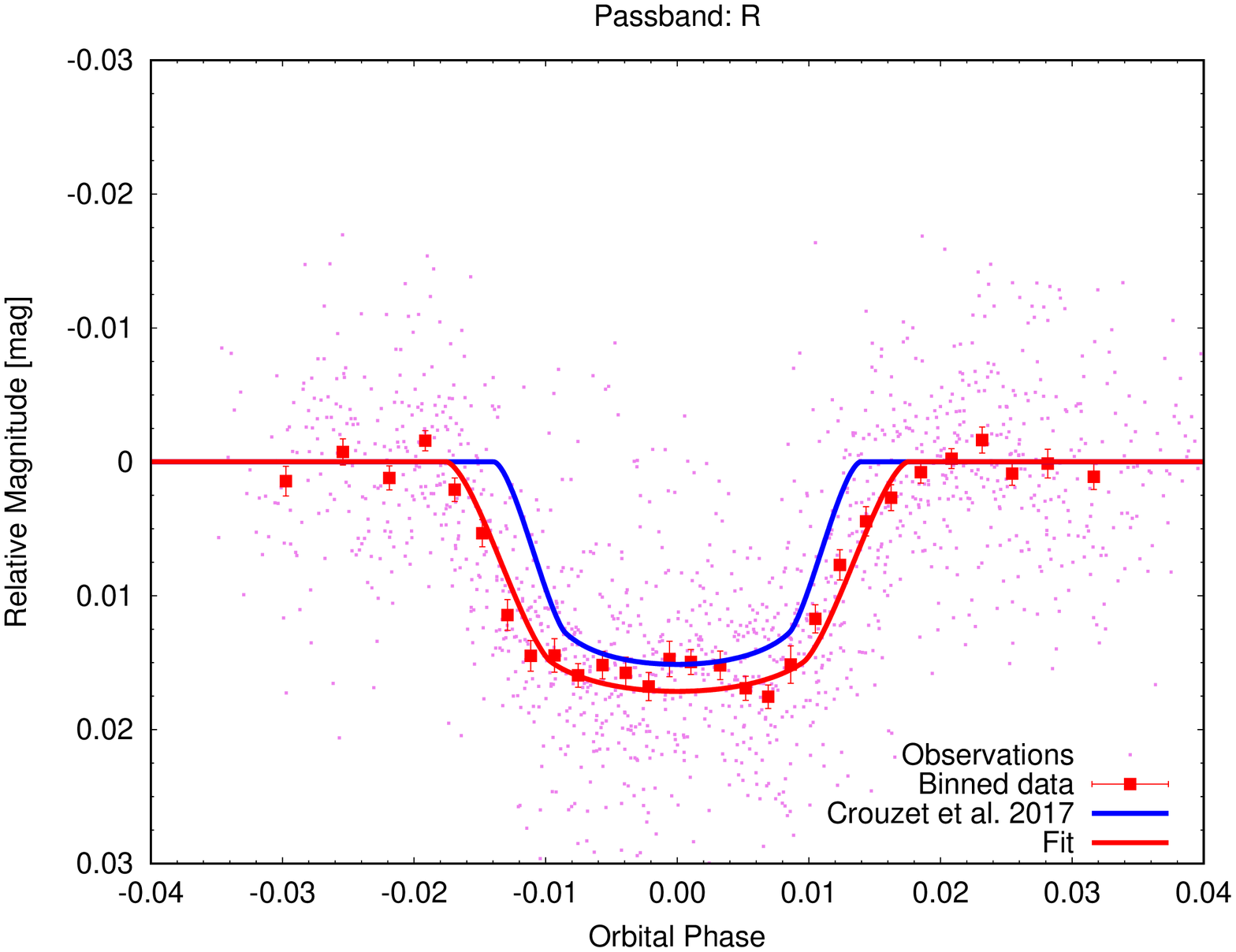}
\includegraphics[width=88mm]{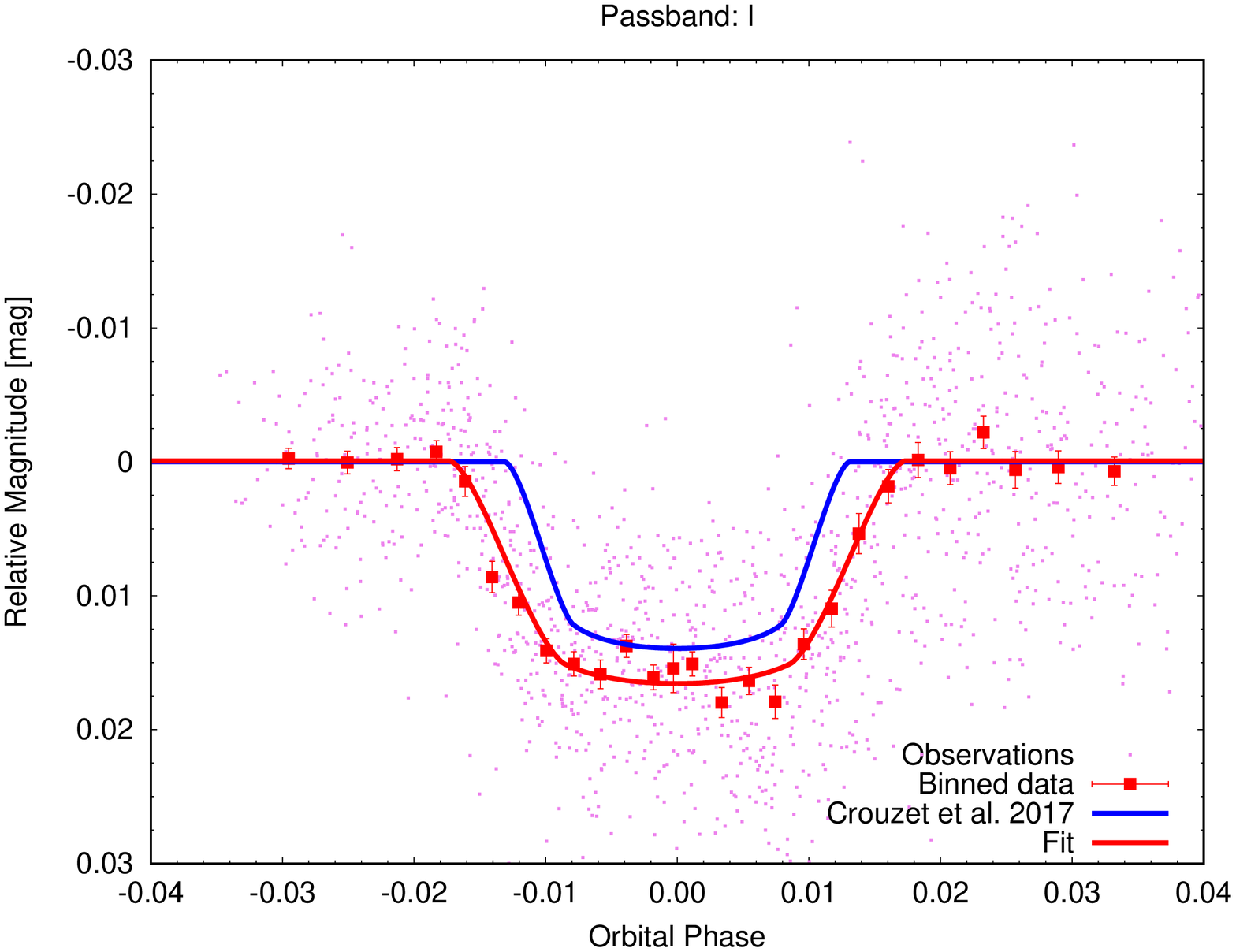}}
\caption{The phase-folded transit light curves of XO-6b in the \textit{B}, \textit{V}, \textit{R}, and \textit{I} passbands for all nights and telescopes combined (Observations), overplotted with the {\tt{JKTEBOP}} best-fit model (red lines). We binned the data points only for better visualisation of the transit shape, but we fitted individual data points. The reproduced best-fit models, obtained by C17, are also shown (blue lines).}
\label{fitsBVRI}
\end{figure*}

The {\tt{JKTEBOP}} best-fit parameters calculated for the \textit{B}, \textit{V}, \textit{R}, and \textit{I} passbands are summarized in Table \ref{paramperbands} separately and the corresponding phase-folded transit light curves of XO-6b, overplotted with the {\tt{JKTEBOP}} best-fit models are shown in Fig. \ref{fitsBVRI}. \citet{Crouzet1} published 3 parameters obtained in \textit{B}, \textit{V}, \textit{R}, and \textit{I} passbands: the normalized semi-major axis $a/R_\mathrm{s}$, the planet-to-star radius ratio $R_\mathrm{p}/R_\mathrm{s}$, and the orbit inclination angle $i$. Since the {\tt{JKTEBOP}} code calculates the sum of fractional radii $(R_\mathrm{p} + R_\mathrm{s})/a$ instead of the normalized semi-major axis $a/R_\mathrm{s}$, we cannot directly compare these parameter values. On the other hand, we can compare the planet-to-star radius ratio and the orbit inclination angle. Table \ref{paramperbands} shows the discussed parameter values, obtained by C17, as well. We can see that the precision of the parameter values is similar. If we compare parameter values obtained for $R_\mathrm{p}/R_\mathrm{s}$ and for $i$, presented by \citet{Crouzet1} and from this work, we can conclude that there is more than $3\sigma$ difference in the case of the planet-to-star radius ratio in the \textit{R} and \textit{I} passbans, and in the case of the orbit inclination angle in all passbands. Based on our photometric data, the planet XO-6b seems to be larger in the \textit{R} and \textit{I} passbands and its orbit inclination angle seems to be smaller than it was determined originally by C17. 

Since this difference is interesting, we checked our results by reproducing the best-fit models, obtained by \citet{Crouzet1}. For this purpose we used the corresponding best-fit parameters, published by C17. The parameter $(R_\mathrm{p} + R_\mathrm{s})/a$ was calculated from the parameters $a/R_\mathrm{s}$ and $R_\mathrm{p}/R_\mathrm{s}$, and the quadratic LD coefficients were calculated based on the stellar parameters of $T_\mathrm{eff} = 6720$ K, log $g = 4.036$ (cgs), m/H $= 0.0$, and $V_\mathrm{micro} = $ 2 km~s$^{-1}$, using the {\tt{JKTLD}}\footnote{See \url{https://www.astro.keele.ac.uk/jkt/codes/jktld.html}.} code \citep{Southworth2} identically as per \citet{Crouzet1}. We reproduced the best-fit models of C17 with the {\tt{JKTEBOP}}-task No. 2, which calculates a synthetic light curve (10~000 points between phases 0 and 1) using the parameters in the input file. These synthetic light curves were re-plotted onto the Fig. \ref{fitsBVRI}. We can clearly see that these best-fit models do not fit our new observations properly: in the cases of passbands \textit{B} and \textit{V}, only the transit depth is fitted well, while in the cases of passbands \textit{R} and \textit{I}, the fit is completely inadequate, which is in agreement with the more than $3\sigma$ difference in the parameter $R_\mathrm{p}/R_\mathrm{s}$ in these passbands. 

\begin{figure}
\centering
\includegraphics[width=70mm]{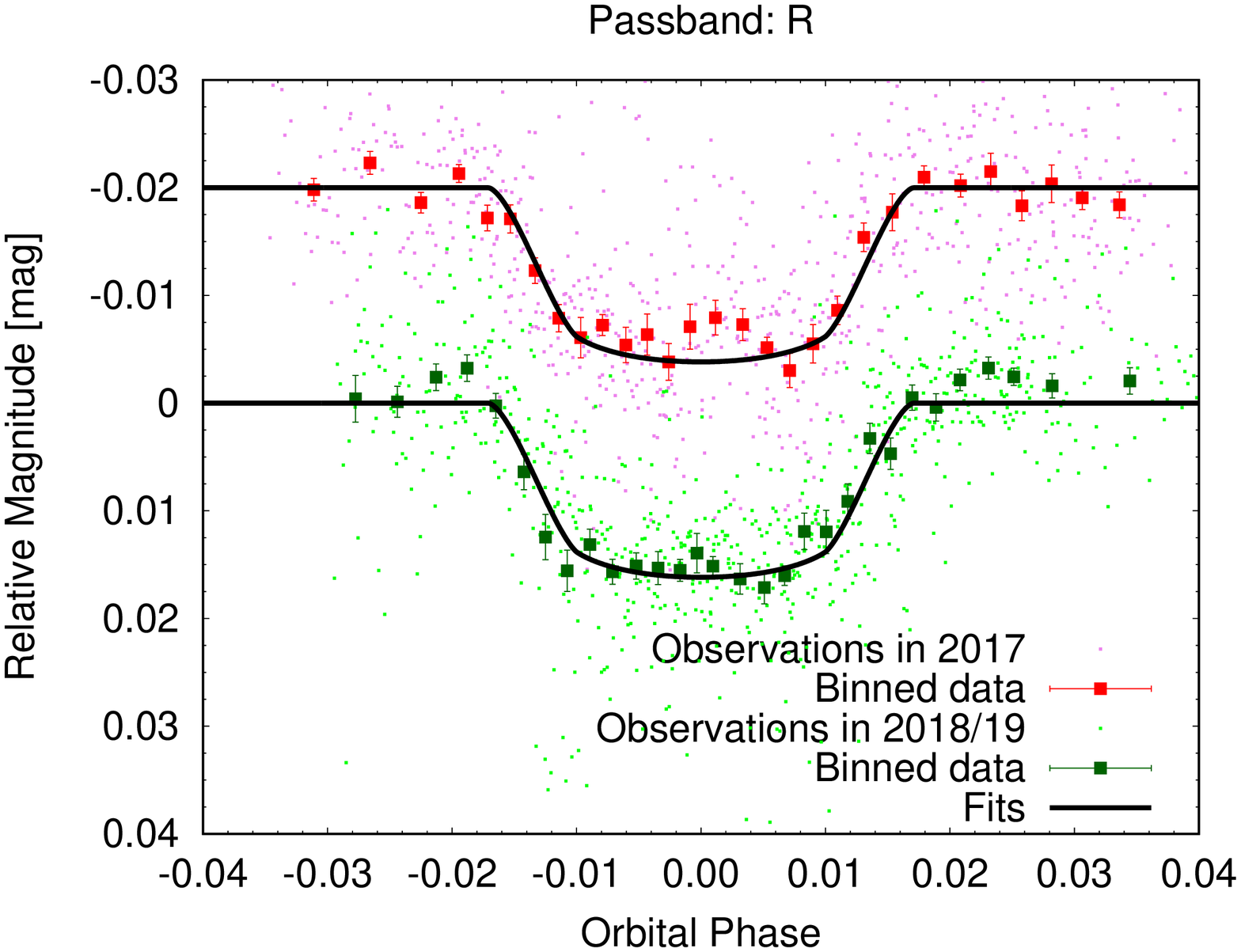}
\caption{Photometric observations of XO-6b transits in the \textit{R} passband, splitted into 2 datasets, overplotted with the {\tt{JKTEBOP}} best-fit model (black lines). We binned the data points only for better visualisation of the transit shape, but we fitted individual data points. The graph shows that there is no evidence for long-term TDVs, nor evidence for a long-term change in the orbit inclination angle $i$.}
\label{splitRfit}
\end{figure}

The more than $3\sigma$ difference in the orbit inclination angle can be due to a long-term change in this parameter and it can be related to TTVs in the system (See Sect. \ref{ttv}). We tested the possibility of a long-term change in the orbit inclination angle by splitting our observations in \textit{R} passband into 2 parts. The first part contains our observations from the year 2017 and the second part our remaining observations (See Table \ref{photobslog}). There is a time-gap of about 3/4 year between the first and the second dataset. We fitted these datasets independently using the {\tt{JKTEBOP}} code (Fig. \ref{splitRfit}). During the fitting procedure only the orbit inclination angle parameter $i$ was freely adjusted, other parameters were fixed to the best values, obtained during the previous fitting procedure in \textit{R} passband (See Table \ref{paramperbands}). As a result, we obtained $i = 84.8(4)$ and $i = 84.6(2)$ deg for the first and the second dataset, respectively, which means that there is no long-term change in the orbit inclination angle, or it needs more precise photometric data, or it appears on a longer time scale only. Since our observations were collected during 2 years, such a time scale can be more than 2 years. Observations of C17 and our observations together were obtained during 5 years, so we cannot rule out the possibility that the mentioned difference in the orbit inclination angle derived in this work and by C17 due to a long-term change. Another possibility is a degeneracy between the parameters $R_\mathrm{p}/R_\mathrm{s}$ and $i$. Since C17 derived a smaller planet size with a larger orbit inclination value, and we got a larger planet size with a smaller orbit inclination value (see and compare corresponding weighted average values in Table \ref{paramperbands}), this scenario is also possible. During this analysis, based on the mentioned 2 datasets in \textit{R} passband, we found no evidence for long-term Transit Duration Variations (TDVs).    

\subsection{Simultaneous analysis of transits and RV measurements}
\label{trandrvanalysis}

As a next step, we fitted all observed transits and RV measurements simultaneously. Since the {\tt{JKTEBOP}} code cannot fit light curves in 4 passbands simultaneously with RV measurements, we developed an own code, called {\tt{RMF}} (Roche ModiFied), where the fitted parameters are simultaneously calculated from multicolor photometric and spectroscopic data. The software was prepared based on the {\tt{ROCHE}} code, which is devoted to the modeling of multi-dataset observations of close eclipsing binary stars, such as radial velocities, multicolor light curves, and broadening functions \citep{Pribulla2, Pribulla3}. During this analysis we used all 8 transits in all passbands simultaneously, our 7 RV measurements, and we also adopted 13 RV values presented by \citet{Crouzet1}, which were obtained outside the transit, to better cover the orbital phase with observations. The latter RV observations were obtained between September 2013 and January 2016 with the SOPHIE spectrograph at the Observatoire de Haute-Provence (OHP) in France. SOPHIE is a cross-dispersed, fibre-fed echelle spectrograph, mounted on the 1.93 m telescope at OHP, which can operate in two observation modes: in high-efficiency mode ($R = 39~000$) to favor high throughput, or in high-resolution mode ($R = 75~000$) to favor better precision. SOPHIE covers the spectral range 3870 -- 6940 \AA~ \citep{Perruchot1, Bouchy1}. Since the RV measurements, published by C17, had an offset in comparison with our RV values (See Sect. \ref{ttv} and Fig. \ref{systemicrv}), we shifted systemic radial velocities $\gamma$ to zero km~s$^{-1}$, by adjusting only the systemic RV parameter in the {\tt{RMF}} code. Up to 3 iteration steps were sufficient for our purposes. In this way we could combine our and the adopted RV values. An overview of the shifted 20 RV values are presented in Table \ref{spectroalllog}.

\begin{table}
\centering
\caption{Log of shifted RV observations, used in our analysis. RV values marked with a $*$ were adopted from \citet{Crouzet1}. Table shows RV values and their $\pm 1\sigma$ uncertainties, sorted by the BJD.}
\label{spectroalllog}
\begin{tabular}{ccc}
\hline
\hline
BJD$ - 2~400~000$ &    RV values [km~s$^{-1}$]  &  $\pm 1\sigma$ [km~s$^{-1}$]\\
\hline
$56551.65530$ &  $-0.49$$^*$ 	&  $0.10$\\
$56560.65920$ &  $+0.07$$^*$ 	&  $0.11$\\
$56560.66170$ &  $-0.10$$^*$ 	&  $0.10$\\
$56723.38580$ &  $+0.38$$^*$ 	&  $0.08$\\
$56974.54510$ &  $-0.17$$^*$ 	&  $0.09$\\
$57363.62540$ &  $+0.04$$^*$ 	&  $0.08$\\
$57378.49220$ &  $+0.22$$^*$ 	&  $0.08$\\
$57383.47300$ &  $-0.08$$^*$ 	&  $0.09$\\
$57384.53890$ &  $+0.15$$^*$ 	&  $0.08$\\
$57399.37420$ &  $-0.13$$^*$ 	&  $0.11$\\
$57402.39770$ &  $-0.19$$^*$ 	&  $0.08$\\
$57402.67580$ &  $-0.10$$^*$ 	&  $0.09$\\
$57405.51910$ &  $+0.04$$^*$ 	&  $0.10$\\
$58374.55832$ &  $-0.90$     	&  $0.50$\\
$58381.60650$ &  $-0.90$     	&  $0.60$\\
$58382.55065$ &  $-1.10$     	&  $0.60$\\
$58530.30396$ &  $-0.30$     	&  $0.80$\\
$58531.34452$ &  $+0.50$     	&  $0.70$\\
$58532.34586$ &  $+0.00$     	&  $0.60$\\
$58533.34734$ &  $+0.30$     	&  $0.60$\\
\hline
\hline
\end{tabular}
\end{table}   

\begin{figure}
\centering
\includegraphics[width=70mm]{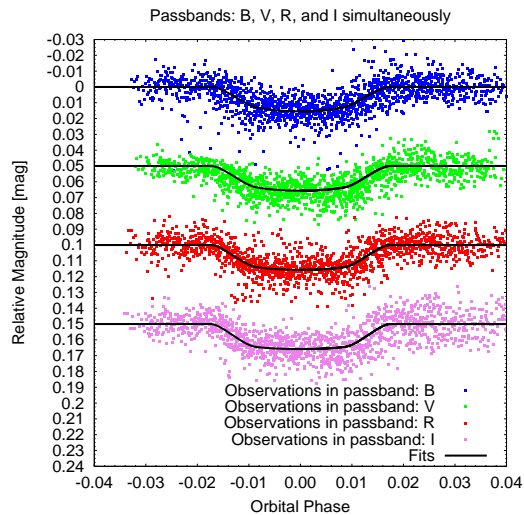}
\caption{The phase-folded transit light curves of XO-6b in the \textit{B}, \textit{V}, \textit{R}, and \textit{I} passbands for all nights and telescopes combined, overplotted with the {\tt{RMF}} best-fit model (black lines). The model was calculated based on simultaneous fit to the all photometric and RV data.}
\label{simfitBVRI}
\end{figure}

\begin{figure}
\centering
\includegraphics[width=70mm]{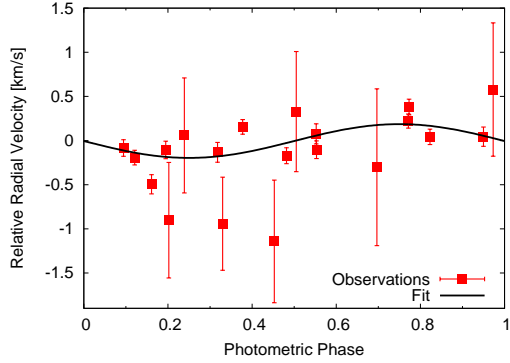}
\caption{The shifted RV measurements of XO-6 (Observations), phased with the orbital period of $P_\mathrm{orb} = 3.765009$ days, and overplotted with the {\tt{RMF}} best-fit model (black line). The model was calculated based on simultaneous fit to the all photometric and RV data.}
\label{rvplanet}
\end{figure}

During the fitting procedure, we freely adjusted the same 6 free parameters as in Sect. \ref{indtransitanalysis}, except the sum of fractional radii, because the code {\tt{RMF}} uses the normalized star radius $R_\mathrm{s}/a$, instead of $(R_\mathrm{p} + R_\mathrm{s})/a$. In addition, for the spectroscopic measurements, we set free the RV semi-amplitude $K$ and the systemic radial velocity $\gamma$. Based on the results of C17, we assumed circular orbit, therefore we fixed the numerical eccentricity $e$ to zero and the periastron longitude $\omega$ to 90 deg. The LD coefficients were also fixed during the fitting procedure, and in this case we used the 4-parameter LD formula and the critical-foreshortening-angle approach \citep{Claret2}. The coefficients ($a_1$, $a_2$, $a_3$, $a_4$, and $\mu_\mathrm{cri}$) were calculated for the \textit{B}, \textit{V}, \textit{R}, and \textit{I} passbands by A. Claret (personal communication\footnote{E-mail: \url{claret@iaa.es}.}), using the same spherical {\tt{PHOENIX-COND}} models as in \citet{Claret2}. Subsequently, we selected the coefficients for the stellar parameters of $T_\mathrm{eff} = 6700$ K, log $g = 4.0$ (cgs), and m/H $= 0.0$, which very well correspond to the parent star XO-6. The uncertainties in the fitted parameters were determined based on covariance matrix. The {\tt{RMF}} best-fit parameters are comparatively summarized in Table \ref{paramallbands}. The corresponding phase-folded transit light curves of XO-6b, overplotted with the {\tt{RMF}} best-fit models are shown in Fig. \ref{simfitBVRI}, and the RV measurements of XO-6 with the best-fit spectroscopic-orbit model is presented in Fig. \ref{rvplanet}. 

From Table \ref{paramallbands} we can see that in comparison with C17, we derived the parameter values with better accuracy in the cases of the planet-to-star radius ratio $R_\mathrm{p}/R_\mathrm{s}$, orbital period of the planet $P_\mathrm{orb}$, and in the case of the RV semi-amplitude $K$. These parameter values are in good agreement within $3\sigma$, including the parameter values for $R_\mathrm{p}/R_\mathrm{s}$, previously discrepant in the \textit{R} and \textit{I} passbands (See Sect. \ref{indtransitanalysis}). The reason is not a radical change in our best-fit estimate, but the relatively large uncertainty in the parameter $R_\mathrm{p}/R_\mathrm{s}$, obtained by C17 (in comparison with the values presented in Table \ref{paramperbands}). Based on our measurements, the planet XO-6b seems to be about 10\% larger, which is, however, only about $2\sigma$ difference. Furthermore, we derived the orbit inclination angle $i$ with the same uncertainty as per \citet{Crouzet1}. On the other hand, we can calculate that there is about $9.5\sigma$ difference between our derived value of $i$, and the value of $i$ presented in the discovery paper. This joint solution confirmed our previous results about the orbit inclination, obtained during the analysis of transits per passband -- we can conclude again that based on our data, the orbit inclination angle of the planet XO-6b seems to be smaller than it was determined originally by C17. This difference can be due to a long-term change in the orbit inclination angle, visible on a longer time scale only, or, alternatively, it can be due to a degeneracy between the parameters $R_\mathrm{p}/R_\mathrm{s}$ and $i$. Another more than $3\sigma$ disagreement is in the mid-transit time parameter values. Our derived value of $T_\mathrm{c}$ has bigger uncertainty, but this can be due to the TTVs (See Sect. \ref{ttv}). The derived orbital period of the planet can be also influenced by this effect. 

\begin{table}
\centering
\caption{An overview of {\tt{RMF}} best-fit parameters simultaneosly calculated from photometric (\textit{B}, \textit{V}, \textit{R}, and \textit{I} passbands), and spectroscopic data. Fixed parameters are listed without uncertainties. Parameter values presented by \citet{Crouzet1} were listed only for easier comparison with the corresponding values from this work.}
\label{paramallbands}
\begin{tabular}{ccc}
\hline
\hline
Parameters 				&	\citet{Crouzet1} &	This work\\
\hline
$R_\mathrm{s}/a$		 	&	--		 &	$0.132(3)$\\
$a/R_\mathrm{s}$		 	&	$9.08(17)$	 &	--\\
$R_\mathrm{p}/R_\mathrm{s}$ 		&	$0.110(6)$	 &	$0.1224(16)$\\
$i$ [deg]				&	$86.0(2)$	 &	$84.1(2)$\\
$e$					&	$0.0$		 &	$0.0$\\	
$\omega$ [deg]				&	$90.0$		 &	$90.0$\\	
$P_\mathrm{orb}$ [days]			&	$3.765000(8)$	 &	$3.765009(4)$\\ 
$T_\mathrm{c} - 2~456~652$ [BJD]	&	$0.7124(5)$ 	 &	$0.706(2)$\\
$L_\mathrm{sf}$ [mag]			&	--		 &	$-0.00093(16)$\\
$K$ [km~s$^{-1}$]			&	$0.20(7)$	 &	$0.19(3)$\\	
$\gamma$ [km~s$^{-1}$]			&	--		 &	$0.00(2)$\\	
\hline
\hline
\end{tabular}
\end{table}   

\section{Transit timing variations of XO-6b}
\label{ttv}

Transit timing observations turned out to be a crucial tool in the study of systems with known multiple transiting planets, see for example \citet{Holman1, Grimm1} and references therein. Periodic TTVs can also indicate a third object in the system, which can be a stellar object, or a brown dwarf, as well as a transiting, or a non-transiting planet, see for example \citet{Holman2, Cochran1, Ballard1, Nesvorny1, Nesvorny2, Fabrycky1, Ofir1}. Although \citet{Crouzet1} found no evidence for TTVs, we decided to investigate the possibility of changes in the timings of XO-6b transits. We were motivated by the O-C diagram, presented in the ETD, which shows clear TTVs\footnote{See \url{http://var2.astro.cz/ETD/etd.php?STARNAME=XO-6&PLANET=b}.}. Based on the data, presented in the ETD up to November 2019, there is a possibility that TTVs of XO-6b can be periodic as well, which means that they can be also due to a third object in the system. As a first step to investigate the possibility of the existence of a third object in the system of XO-6, we constructed an own O-C diagram. We used 15 transits in total: besides our 8 individual transits, we adopted 7 transit observations from the ETD. The main criteria for transit selection from the ETD were (1) the clear indication of the used time-stamp of observations as JD, or HJD, and (2) the quality of the photometric measurement. Every JD, or HJD time-stamp was then converted to the BJD$_\mathrm{TDB}$, using the on-line applets {\tt{UTC2BJD}}\footnote{See \url{http://astroutils.astronomy.ohio-state.edu/time/utc2bjd.html}.}, or {\tt{HJD2BJD}}\footnote{See \url{http://astroutils.astronomy.ohio-state.edu/time/hjd2bjd.html}.}, respectively. During the next step, the linear trend, due to the second-order extinction, was removed from the ETD transit light curves and, finally, the outlier data points were removed. We used a $3\sigma$ clipping (where $\sigma$ is the standard deviation of the light curve), as in the case of our photometric data (See Sect. \ref{phot}). The resulting ETD light curves were used in the next analysis.

As a next step we calculated the "C" ("Calculated") times of the mid-transits. This is the first element to construct the O-C diagram. We used the linear ephemeris formula as 

\begin{equation}
\label{linef}
T_0 = T_\mathrm{c} + P_\mathrm{orb} \times E = 2~456~652.706(2) + 3.765009(4) \times E,
\end{equation}
   
\begin{figure}
\centering
\includegraphics[width=70mm]{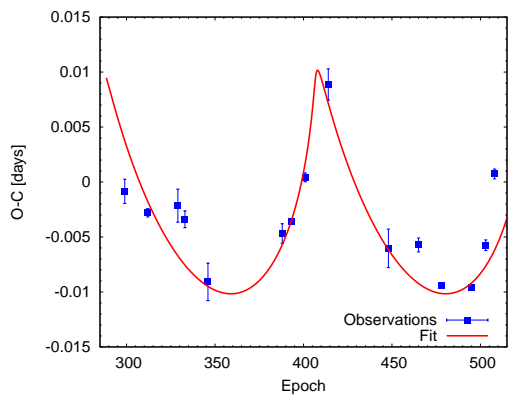}
\caption{The O-C diagram of 15 XO-6b transits. The red line represents the {\tt{OCFit}} best fit computed from the LiTE model.}
\label{fittedocdiagram}
\end{figure}

\noindent{were $T_0$ corresponds to the "C" value, and $E$ is the epoch of observation, i.e. the number of the orbital cycle from $T_\mathrm{c}$. From Table \ref{paramallbands} we used the improved parameter values for $T_\mathrm{c}$ and $P_\mathrm{orb}$ to calculate $T_0$ values. To determine the "O" ("Observed") times of the mid-transits, we individually fitted every transit event using the {\tt{RMF}} code, but during this fitting procedure we fixed every parameter to its best value (See Table \ref{paramallbands}, column "This work"), and we freely adjusted only the mid-transit time parameter, i.e. the calculated "C" value of the selected transit event. The best-fit value of this parameter is the wanted "O" value. In such a way we could construct our own O-C diagram. Uncertainties of individual O-C values are propagated from uncertainties in the parameters $T_\mathrm{c}$ and $P_\mathrm{orb}$ (See Eq. \ref{linef}), and from uncertainties of "O" values. The final O-C dataset is summarized in Table \ref{inputocdata} and the corresponding O-C diagram is plotted in Fig. \ref{fittedocdiagram}. This diagram confirms the TTVs, presented in the ETD, moreover, it seems to be periodic, as well, which indicates the presence of an additional object in the system.}                  

\begin{table}
\centering
\caption{Log of O-C data used in our analysis. O-C values with their $\pm 1\sigma$ uncertainties are sorted by the epoch. The O-C values marked with a $*$ were calculated based on the adopted ETD transits.}
\label{inputocdata}
\begin{tabular}{clc}
\hline
\hline
Epoch 	 &  O-C [days] 		  & $\pm 1 \sigma$ [days]\\
\hline
$299.0$  &  $-0.000800$$^*$	  &  $0.001100$\\
$312.0$  &  $-0.002700$		  &  $0.000300$\\
$329.0$  &  $-0.002100$   	  &  $0.001500$\\
$333.0$  &  $-0.003300$	  	  &  $0.000700$\\
$346.0$  &  $-0.009000$		  &  $0.001700$\\
$388.0$  &  $-0.004600$$^*$	  &  $0.000800$\\
$393.0$  &  $-0.003557$$^*$	  &  $0.000019$\\
$401.0$  &  $+0.000400$$^*$       &  $0.000400$\\
$414.0$  &  $+0.008800$           &  $0.001400$\\
$448.0$  &  $-0.006000$           &  $0.001700$\\
$465.0$  &  $-0.005700$           &  $0.000600$\\
$478.0$  &  $-0.009400$$^*$       &  $0.000190$\\
$495.0$  &  $-0.009620$$^*$       &  $0.000080$\\
$503.0$  &  $-0.005700$$^*$       &  $0.000400$\\
$508.0$  &  $+0.000700$           &  $0.000400$\\
\hline
\hline
\end{tabular}
\end{table}   

There are two plausible explanations of such TTVs. (1) A third object in the system causing light-time effect (LiTE). The parent star -- transiting planet subsystem moves around the common barycentre of a wider triple system. It produces periodic variations in the observed transit times with respect to the linear ephemeris of this system with a period corresponding to the period of the third object \citep{Irwin1}. (2) Another plausible explanation of periodic TTVs could be resonant perturbations between the transiting planet and another unknown low-mass planet in the system. Resonant interactions, mainly mean-motion resonances 1:2 and 2:3, are frequent among exoplanetary systems \citep{Wang2}. To investigate, which explanation is more realistic, we first fitted the O-C values using the {\tt{OCFit}} code \citep{Gajdos1}. It is a relatively new software, but it was already used in \citet{Gajdos4, Gajdos3, Gajdos2} with success. The package uses Genetic Algorithms (GA) and Markov Chain Monte Carlo (MCMC) methods. Unlike many others fitting routines, this routine does not need any starting values of fitted parameters, only starting intervals. These intervals can be quite large. Fitting using the software is simple thanks to a very intuitive graphic user interface. Currently, 9 most common models of periodic O-C changes are included in this software. From these offered models we used the LiTE model, which is based on the light-time effect.

As first, we used the GA method to calculate the initial parameter values. These values together with the estimation of their uncertainties were finalized using the MCMC method. To get final parameter values we used $10^6$ iteration steps. The {\tt{OCFit}} best-fit parameters are summarized in Table \ref{ocfitparams} and the O-C diagram, overplotted with the best-fit LiTE model is shown in Fig. \ref{fittedocdiagram}.This solution assumes a third object in the system, which causes mid-transit time-shifts in XO-6b transits of about $\pm 14$ min around the zero value. This is the semi-amplitude $K_3$, and it is clearly seen in the O-C diagram. The orbital period of the predicted third object ($P_\mathrm{orb3}$) is about 450 days. In Fig. \ref{fittedocdiagram}, we can see that the LiTE model fits the observations relatively well, the $\chi^\mathrm{2}$, or the $\chi^\mathrm{2}_\mathrm{red}$ value, which is the goodness-of-fit parameter, is about 400 and 40, respectively. On the other hand, the model requires quite eccentric orbit for the assumed third object ($e \approx 0.8$, with the pericenter longitude of $\omega_3 \approx 53$ deg). The model predicts the third object with a stellar mass. Since the orbit inclination of this object $i_3$ is unknown, using the LiTE model we can calculate only its mass-function $f(M_\mathrm{3})$ as 

\begin{equation}
\label{linef2}
f(M_\mathrm{3}) = \frac{(M_3~\sin~i_3)^3}{M^2} = \frac{(a~\sin~i_3)^3}{P_\mathrm{orb3}^2} = 5.3(5)~\mathrm{M}_\odot,
\end{equation}            

\noindent{where $M_3$ is the true mass of the third object, $a$ is the semi-major axis of the orbit of XO-6, and $M$ is the total mass of the system. Finally, the program also calculates the time of pericenter passage of the predicted third object $T(0)_3$, which is $2~458~184(2)$ [BJD].} 

\begin{table}
\centering
\caption{An overview of {\tt{OCFit}} best-fit parameters of the assumed third object in the system XO-6, obtained from the LiTE model. For more details see the text in Sect. \ref{ttv}.}
\label{ocfitparams}
\begin{tabular}{cc}
\hline
\hline
Parameters 			& LiTE model      \\
\hline
$P_\mathrm{orb3}$ [days]    	&  $456(3)$	  \\
$P_\mathrm{orb3}$ [years]   	&  $1.251(8)$	  \\
$a~\sin~i_3$ [AU]	  	&  $2.03(6)$  	  \\
$e_3$			    	&  $0.85(2)$  	  \\
$T(0)_3$ [BJD]		    	&  $2458184(2)$   \\
$\omega_3$ [deg]	   	&  $53(3)$   	  \\
$K_3$ [sec]	 	   	&  $878(38)$      \\
$K_3$ [min]	 	   	&  $14.6(6)$      \\
$f(M_\mathrm{3})$ [M$_\odot$] 	&  $5.3(5)$       \\
\hline
$\chi^2$			&  $406.08$	  \\
$\chi^2_\mathrm{red}$		&  $40.60$        \\
\hline
\hline
\end{tabular}
\end{table}   

\begin{figure*}
\centering
\centerline{
\includegraphics[width=70mm]{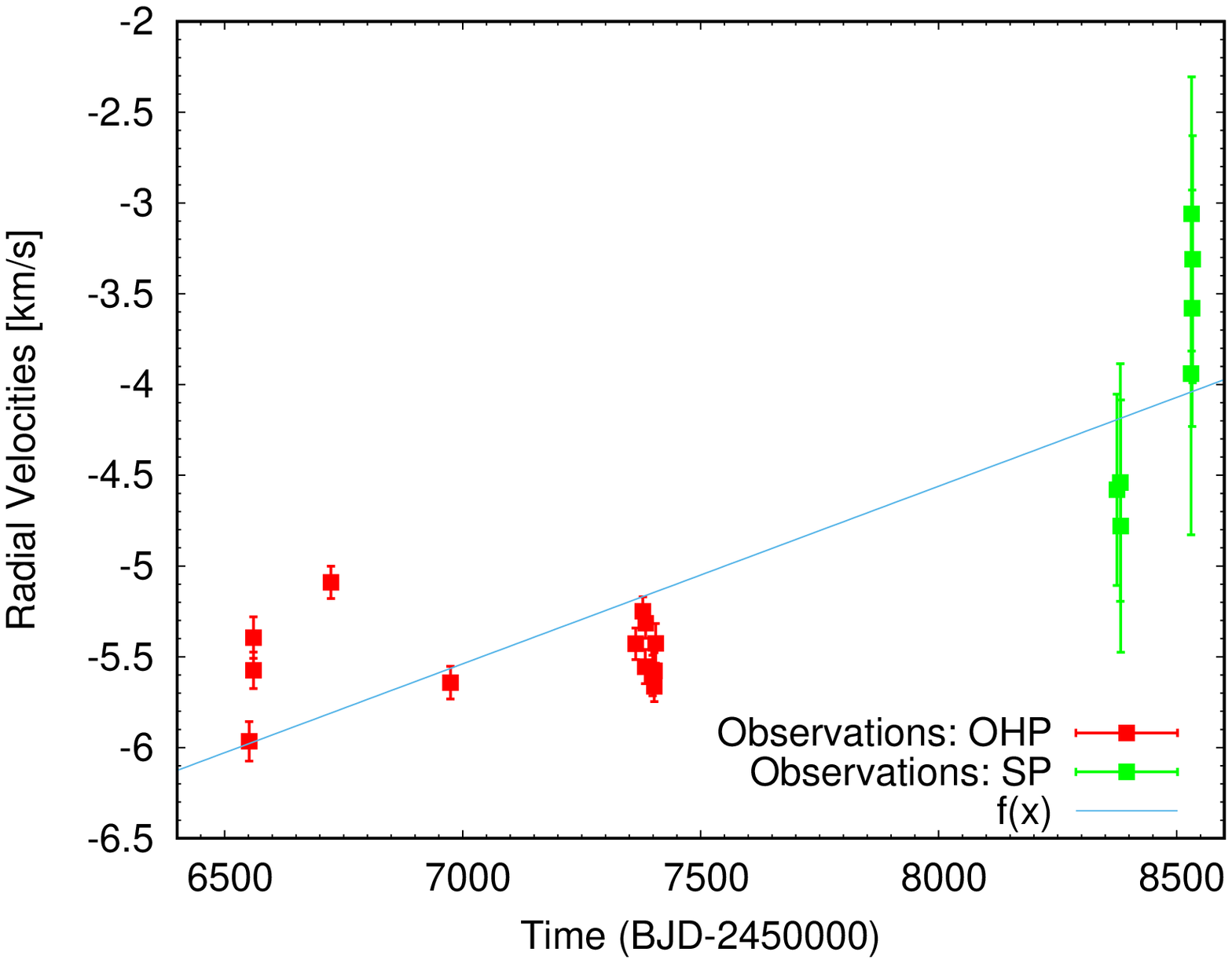}
\includegraphics[width=70mm]{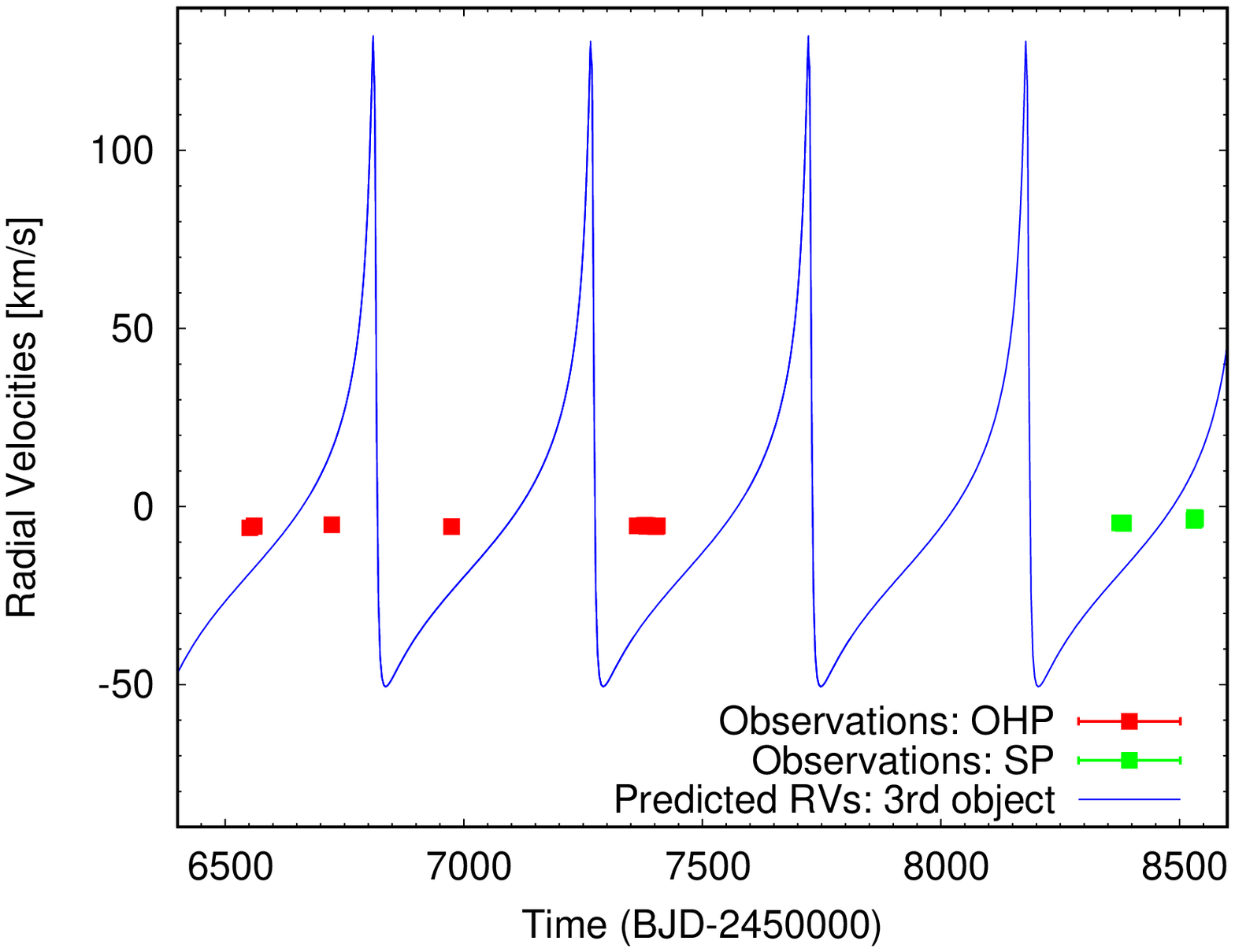}}
\caption{The RV values used in our analysis. The red points represent the 13 RV measurements adopted from \citet{Crouzet1}, while the green points are our 7 RV values. The RV measurements were fitted with a linear function to better visualize the offset between the two datasets (left panel). The RV measurements are compared with the spectroscopic orbit of the assumed third object in the system XO-6, computed from the LiTE model using the {\tt{RMF}} code (right panel).}
\label{systemicrv}
\end{figure*}   

Since the LiTE model predicts a third object with a stellar mass, we should observe large amplitude variations of observed RV measurements of XO-6, which we can calculate as

\begin{equation}
\label{linef4}
K_\mathrm{RV3} = \frac{2 \pi a \sin i_3}{P_\mathrm{orb3} \sqrt{1 - e_3^{2}}},
\end{equation}

\noindent{where $K_\mathrm{RV3}$ is the semi-amplitude of the RV motion of the parent star XO-6, induced by the predicted third object in the system. Using the corresponding parameter values derived from the LiTE model (See Table \ref{ocfitparams}) we can obtain the RV semi-amplitude of about 92 km~s$^{-1}$. Now, we can analyze the RVs from the viewpoint of the assumed third object. Previously, we shifted systemic radial velocities $\gamma$ to zero km~s$^{-1}$ (See Sect. \ref{trandrvanalysis}). In this way we could combine our and the RV values adopted from \citet{Crouzet1}. This step was necessary, because we found that RV measurements, published by C17, had an offset in comparison with our RV values. The RV offset is about 2.5 km~s$^{-1}$ as it is clearly seen in Fig. \ref{systemicrv} (left panel). Although this offset is too small in comparison with the prediction, theoretically, it can be related to the assumed third object in the system. To confirm or reject this relation, we tried to find a joint solution of RV measurements and O-C observations using the code {\tt{RMF}}. All attempts to fit both datasets simultaneously failed to provide a consistent solution, which means that very probably there is no relation between the observed RV offset and TTVs. There is no RV signal from the predicted third object, as we can see in Fig. \ref{systemicrv} (right panel). All RVs are distributed close to the value of $-5$ km~s$^{-1}$. Although there are some longer gaps in RV observations, as well as relatively low number of these observations, we should observe the RV motion of XO-6, induced by the third (stellar) object in the system, in different phases, therefore we should obtain RV values from much larger interval, which is not the case. This means that the first explanation, i.e. a third object in the system causing LiTE seems to be unlikely. This theory is supported by the fact that no indication of another star is seen in the spectra, as it was reported also by \citet{Crouzet1}. Therefore, the second explanation seems to be more realistic: resonant perturbations between the transiting planet XO-6b and another unknown low-mass planet in this system.}              

\section{Conclusions}
\label{conc}

We used the telescopes at AI SAS for follow-up observations of the newly discovered transiting exoplanet XO-6b. We started our multicolor photometric and RV observations shortly after the discovery paper was published by \citet{Crouzet1}. Our main scientific goals were to better characterize the planetary system and to search for TTVs. During two years of observations, between March 2017 and March 2019, we obtained 8 photometric transit measurements in \textit{B}, \textit{V}, \textit{R}, and \textit{I} passbands, and 7 RV measurements with our instruments. Unfortunately, we had to discard several measurements due to the low data quality.

As first, we analyzed the observed transits per passband, using the {\tt{JKTEBOP}} code. During the next step, we analyzed all observed transits and RV measurements simultaneously with the code {\tt{RMF}}. Besides our 7 RV measurements, we adopted 13 RV values from C17, which were obtained outside the transit, to better cover the orbital phase with observations. During the second part of our work we investigated the timings of XO-6b transits. As first, we constructed an own O-C diagram of XO-6b transits. We used our 8 transit measurements, and 7 observations were taken from the ETD to increase the number of data points in the diagram. Subsequently, we analyzed the resulting O-C diagram using the LiTE model, included in the {\tt{OCFit}} code. Finally, we tried to find a joint solution of RV measurements and O-C observations using the code {\tt{RMF}}.  

Based on our results we can conclude that the system is quite interesting. Previously, C17 found that the planet XO-6b orbits a relatively bright ($V \approx 10.2$ mag), hot ($T_\mathrm{eff} \approx 6720$ K), and fast rotating ($v~\mathrm{sin}~i \approx 48$ km s$^{-1}$) parent star XO-6 on a prograde and misaligned orbit ($\lambda \approx -20$ deg). Only a few similar exoplanets are known. We refined several planetary and orbital parameters. Based on analysis of transits per passband, we found that our best-fit planet-to-star radius ratio parameter values, $R_\mathrm{p}/R_\mathrm{s} = 0.124(2)$ and $R_\mathrm{p}/R_\mathrm{s} = 0.123(2)$, differ more than $3\sigma$ in comparison with the values presented by C17, $R_\mathrm{p}/R_\mathrm{s} = 0.1153(19)$ and $R_\mathrm{p}/R_\mathrm{s} = 0.1114(17)$, in \textit{R} and \textit{I} passbands, respectively. Based on simultaneous analysis of all observed transits and RV measurements, we found the planet-to-star radius ratio parameter values in good agreement within $3\sigma$, but we derived this parameter with better accuracy ($R_\mathrm{p}/R_\mathrm{s} = 0.1224(16)$) than it was presented in the discovery paper ($R_\mathrm{p}/R_\mathrm{s} = 0.110(6)$). The difference in this case is only about $2\sigma$, which is about 10\%, and if we take $R_\mathrm{s} = 1.86$ R$_\odot$ as a radius of the parent star XO-6, we can get the radius of the planet of $R_\mathrm{p} = 0.227(2)$ R$_\odot$ instead of $R_\mathrm{p} = 0.204(11)$ R$_\odot$. Furthermore, we found a more than $3\sigma$ discrepancy in the orbit inclination angle parameter values. In all passbands we obtained smaller values of $i$ than C17. Based on simultaneous analysis of all observed transits and RV measurements, we found the orbit inclination angle of $i = 84.1(2)$ deg, while \citet{Crouzet1} obtained $i = 86.0(2)$ deg, which is about $9.5\sigma$ difference. We can conclude that based on our data, the orbit inclination angle of the planet XO-6b seems to be smaller than it was determined originally by C17. In addition, we tested the possibility of a long-term change in the orbit inclination angle by splitting our observations in \textit{R} passband into 2 parts, but we found that there is no long-term change in this parameter, or it needs more precise photometric data, or it appears on a longer time scale only (more than 2 years). Another possibility is a degeneracy between the parameters $R_\mathrm{p}/R_\mathrm{s}$ and $i$. Another more than $3\sigma$ disagreement is in the mid-transit time parameter values. This can be due to the TTVs, and the derived orbital period of the planet can be also influenced by this effect. During this analysis we found no evidence for long-term TDVs.          

Finally, we found that the system XO-6 shows TTVs. These variations seems to be periodic with the period of about 450 days, therefore, at the beginning of our analysis we proposed two plausible explanations of such TTVs: (1) a third object in the system causing LiTE, or (2) resonant perturbations between the transiting planet and another unknown low-mass planet in the system. From the O-C diagram we derived that the assumed third object in the system should have a stellar mass, therefore significant variations are expected in the RV measurements of XO-6. We found that RV measurements published by C17 are shifted in comparison with our RV values. The RV offset is about 2.5 km~s$^{-1}$. On the other hand, this offset is too small in comparison with the RV semi-amplitude of about 92 km~s$^{-1}$, corresponding to the LiTE model. Moreover, all attempts to fit RVs and O-C data simultaneously failed to provide a consistent solution. This means that the first explanation of the observed TTVs seems to be unlikely and that more realistic is the second explanation. 

In the near future we would like to perform numerical simulations to study mean-motion resonances with XO-6b and to identify stable regions, where the additional planet could exist for a long time. Recently, the system Kepler-410A was studied in a similar manner by \citet{Gajdos3}, and a small unknown planet with a mass of 1.5 M$_\mathrm{Mars}$ was proposed as a perturber object, which causes the observed TTVs in this system. In the near future, we also plan a high-precision RV campaign, which will be needed to measure the mass of the planet XO-6b with greater confidence. Our RV measurements are not sufficiently precise to derive this parameter. This is mainly due to the fast rotating parent star, which complicates RV measurements.                            

\section*{Acknowledgements}

The authors thank the contributors of the ETD, namely F. Garcia, P. Guerra, F. Scaggiante, D. Zardin, M. Bretton, T. Medulka, S. Gudmundsson, and J. Borel, for photometric observations of XO-6b transits. We also thank V. Koll\'{a}r and P. Sivani\v{c} for the technical assistance. This work was supported by the VEGA grant of the Slovak Academy of Sciences No. 2/0031/18, by the Slovak Research and Development Agency under the contract No. APVV-15-0458, by the realization of the Project ITMS No. 26220120029, based on the Supporting Operational Research and Development Program financed from the European Regional Development Fund, by the Hungarian NKFI grant No. K-119517 and the GINOP No. 2.3.2-15-2016-00003 of the Hungarian National Research, Development and Innovation Office, and by the City of Szombathely under agreement No. 67.177-21/2016. 

%%%%%%%%%%%%%%%%%%%%%%%%%%%%%%%%%%%%%%%%%%%%%%%%%%

%%%%%%%%%%%%%%%%%%%% REFERENCES %%%%%%%%%%%%%%%%%%

% The best way to enter references is to use BibTeX:

\bibliographystyle{mnras}
\bibliography{Yourfile} % if your bibtex file is called example.bib

\begin{thebibliography}{}
\makeatletter
\relax
\def\mn@urlcharsother{\let\do\@makeother \do\$\do\&\do\#\do\^\do\_\do\%\do\~}
\def\mn@doi{\begingroup\mn@urlcharsother \@ifnextchar [ {\mn@doi@}
  {\mn@doi@[]}}
\def\mn@doi@[#1]#2{\def\@tempa{#1}\ifx\@tempa\@empty \href
  {http://dx.doi.org/#2} {doi:#2}\else \href {http://dx.doi.org/#2} {#1}\fi
  \endgroup}
\def\mn@eprint#1#2{\mn@eprint@#1:#2::\@nil}
\def\mn@eprint@arXiv#1{\href {http://arxiv.org/abs/#1} {{\tt arXiv:#1}}}
\def\mn@eprint@dblp#1{\href {http://dblp.uni-trier.de/rec/bibtex/#1.xml}
  {dblp:#1}}
\def\mn@eprint@#1:#2:#3:#4\@nil{\def\@tempa {#1}\def\@tempb {#2}\def\@tempc
  {#3}\ifx \@tempc \@empty \let \@tempc \@tempb \let \@tempb \@tempa \fi \ifx
  \@tempb \@empty \def\@tempb {arXiv}\fi \@ifundefined
  {mn@eprint@\@tempb}{\@tempb:\@tempc}{\expandafter \expandafter \csname
  mn@eprint@\@tempb\endcsname \expandafter{\@tempc}}}

\bibitem[\protect\citeauthoryear{{Bakos} et~al.,}{{Bakos}
  et~al.}{2009}]{Bakos1}
{Bakos} G.~{\'A}.,  et~al., 2009, in {Pont} F.,  {Sasselov} D.,   {Holman}
  M.~J.,  eds,  IAU Symposium Vol. 253, Transiting Planets. pp 21--27,
  \mn@doi{10.1017/S1743921308026197}

\bibitem[\protect\citeauthoryear{{Ballard} et~al.,}{{Ballard}
  et~al.}{2011}]{Ballard1}
{Ballard} S.,  et~al., 2011, \mn@doi [\apj] {10.1088/0004-637X/743/2/200},
  \href {https://ui.adsabs.harvard.edu/abs/2011ApJ...743..200B} {743, 200}

\bibitem[\protect\citeauthoryear{{Baudrand} \& {Bohm}}{{Baudrand} \&
  {Bohm}}{1992}]{Baudrand1}
{Baudrand} J.,  {Bohm} T.,  1992, \aap, \href
  {https://ui.adsabs.harvard.edu/abs/1992A&A...259..711B} {259, 711}

\bibitem[\protect\citeauthoryear{{Bessell}}{{Bessell}}{2005}]{Bessell1}
{Bessell} M.~S.,  2005, \mn@doi [\araa]
  {10.1146/annurev.astro.41.082801.100251}, \href
  {https://ui.adsabs.harvard.edu/abs/2005ARA&A..43..293B} {43, 293}

\bibitem[\protect\citeauthoryear{{Borucki}, {Dunham}, {Koch}, {Cochran},
  {Rose}, {Cullers}, {Granados}  \& {Jenkins}}{{Borucki}
  et~al.}{1996}]{Borucki2}
{Borucki} W.~J.,  {Dunham} E.~W.,  {Koch} D.~G.,  {Cochran} W.~D.,  {Rose}
  J.~D.,  {Cullers} D.~K.,  {Granados} A.,   {Jenkins} J.~M.,  1996, \mn@doi
  [\apss] {10.1007/BF00644220}, \href
  {https://ui.adsabs.harvard.edu/abs/1996Ap&SS.241..111B} {241, 111}

\bibitem[\protect\citeauthoryear{{Borucki} et~al.,}{{Borucki}
  et~al.}{2004}]{Borucki3}
{Borucki} W.,  et~al., 2004, in {Favata} F.,  {Aigrain} S.,   {Wilson} A.,
  eds,  ESA Special Publication Vol. 538, Stellar Structure and Habitable
  Planet Finding. pp 177--182

\bibitem[\protect\citeauthoryear{{Borucki} et~al.,}{{Borucki}
  et~al.}{2011}]{Borucki1}
{Borucki} W.~J.,  et~al., 2011, \mn@doi [\apj] {10.1088/0004-637X/736/1/19},
  \href {https://ui.adsabs.harvard.edu/abs/2011ApJ...736...19B} {736, 19}

\bibitem[\protect\citeauthoryear{{Bouchy} et~al.,}{{Bouchy}
  et~al.}{2009}]{Bouchy1}
{Bouchy} F.,  et~al., 2009, \mn@doi [\aap] {10.1051/0004-6361/200912427}, \href
  {https://ui.adsabs.harvard.edu/abs/2009A&A...505..853B} {505, 853}

\bibitem[\protect\citeauthoryear{{Broeg}, {Fern{\'a}ndez}  \&
  {Neuh{\"a}user}}{{Broeg} et~al.}{2005}]{Broeg2}
{Broeg} C.,  {Fern{\'a}ndez} M.,   {Neuh{\"a}user} R.,  2005, \mn@doi
  [Astronomische Nachrichten] {10.1002/asna.200410350}, \href
  {https://ui.adsabs.harvard.edu/abs/2005AN....326..134B} {326, 134}

\bibitem[\protect\citeauthoryear{{Broeg} et~al.,}{{Broeg}
  et~al.}{2013}]{Broeg1}
{Broeg} C.,  et~al., 2013, in European Physical Journal Web of Conferences. p.
  03005 (\mn@eprint {arXiv} {1305.2270}), \mn@doi{10.1051/epjconf/20134703005}

\bibitem[\protect\citeauthoryear{{Claret}}{{Claret}}{2018}]{Claret2}
{Claret} A.,  2018, \mn@doi [\aap] {10.1051/0004-6361/201833060}, \href
  {https://ui.adsabs.harvard.edu/abs/2018A&A...618A..20C} {618, A20}

\bibitem[\protect\citeauthoryear{{Claret} \& {Bloemen}}{{Claret} \&
  {Bloemen}}{2011}]{Claret1}
{Claret} A.,  {Bloemen} S.,  2011, \mn@doi [\aap]
  {10.1051/0004-6361/201116451}, \href
  {http://adsabs.harvard.edu/abs/2011A%26A...529A..75C} {529, A75}

\bibitem[\protect\citeauthoryear{{Cochran} et~al.,}{{Cochran}
  et~al.}{2011}]{Cochran1}
{Cochran} W.~D.,  et~al., 2011, \mn@doi [\apjs] {10.1088/0067-0049/197/1/7},
  \href {https://ui.adsabs.harvard.edu/abs/2011ApJS..197....7C} {197, 7}

\bibitem[\protect\citeauthoryear{{Crouzet}}{{Crouzet}}{2018}]{Crouzet2}
{Crouzet} N.,  2018, {Small Telescope Exoplanet Transit Surveys: XO}.
p.~129, \mn@doi{10.1007/978-3-319-55333-7_129}

\bibitem[\protect\citeauthoryear{{Crouzet} et~al.,}{{Crouzet}
  et~al.}{2017}]{Crouzet1}
{Crouzet} N.,  et~al., 2017, \mn@doi [\aj] {10.3847/1538-3881/153/3/94}, \href
  {https://ui.adsabs.harvard.edu/abs/2017AJ....153...94C} {153, 94}

\bibitem[\protect\citeauthoryear{{Dawson}, {Johnson}, {Morton}, {Crepp},
  {Fabrycky}, {Murray-Clay}  \& {Howard}}{{Dawson} et~al.}{2012}]{Dawson1}
{Dawson} R.~I.,  {Johnson} J.~A.,  {Morton} T.~D.,  {Crepp} J.~R.,  {Fabrycky}
  D.~C.,  {Murray-Clay} R.~A.,   {Howard} A.~W.,  2012, \mn@doi [\apj]
  {10.1088/0004-637X/761/2/163}, \href
  {https://ui.adsabs.harvard.edu/abs/2012ApJ...761..163D} {761, 163}

\bibitem[\protect\citeauthoryear{{Eastman}, {Siverd}  \& {Gaudi}}{{Eastman}
  et~al.}{2010}]{Eastman2}
{Eastman} J.,  {Siverd} R.,   {Gaudi} B.~S.,  2010, \mn@doi [\pasp]
  {10.1086/655938}, \href
  {https://ui.adsabs.harvard.edu/abs/2010PASP..122..935E} {122, 935}

\bibitem[\protect\citeauthoryear{{Eastman}, {Gaudi}  \& {Agol}}{{Eastman}
  et~al.}{2013}]{Eastman1}
{Eastman} J.,  {Gaudi} B.~S.,   {Agol} E.,  2013, \mn@doi [\pasp]
  {10.1086/669497}, \href {http://adsabs.harvard.edu/abs/2013PASP..125...83E}
  {125, 83}

\bibitem[\protect\citeauthoryear{{Fabrycky} et~al.,}{{Fabrycky}
  et~al.}{2012}]{Fabrycky1}
{Fabrycky} D.~C.,  et~al., 2012, \mn@doi [\apj] {10.1088/0004-637X/750/2/114},
  \href {https://ui.adsabs.harvard.edu/abs/2012ApJ...750..114F} {750, 114}

\bibitem[\protect\citeauthoryear{{Gajdo{\v{s}}} \& {Parimucha}}{{Gajdo{\v{s}}}
  \& {Parimucha}}{2019}]{Gajdos1}
{Gajdo{\v{s}}} P.,  {Parimucha} {\v{S}}.,  2019, Open European Journal on
  Variable Stars, \href {https://ui.adsabs.harvard.edu/abs/2019OEJV..197...71G}
  {197, 71}

\bibitem[\protect\citeauthoryear{{Gajdo{\v{s}}}, {Parimucha}, {Hamb{\'a}lek}
  \& {Va{\v{n}}ko}}{{Gajdo{\v{s}}} et~al.}{2017}]{Gajdos4}
{Gajdo{\v{s}}} P.,  {Parimucha} {\v{S}}.,  {Hamb{\'a}lek} {\v{L}}.,
  {Va{\v{n}}ko} M.,  2017, \mn@doi [\mnras] {10.1093/mnras/stx963}, \href
  {https://ui.adsabs.harvard.edu/abs/2017MNRAS.469.2907G} {469, 2907}

\bibitem[\protect\citeauthoryear{{Gajdo{\v{s}}} et~al.,}{{Gajdo{\v{s}}}
  et~al.}{2019a}]{Gajdos3}
{Gajdo{\v{s}}} P.,  et~al., 2019a, \mn@doi [\mnras] {10.1093/mnras/stz305},
  \href {https://ui.adsabs.harvard.edu/abs/2019MNRAS.484.4352G} {484, 4352}

\bibitem[\protect\citeauthoryear{{Gajdo{\v{s}}} et~al.,}{{Gajdo{\v{s}}}
  et~al.}{2019b}]{Gajdos2}
{Gajdo{\v{s}}} P.,  et~al., 2019b, \mn@doi [\mnras] {10.1093/mnras/stz676},
  \href {https://ui.adsabs.harvard.edu/abs/2019MNRAS.485.3580G} {485, 3580}

\bibitem[\protect\citeauthoryear{{Garai} et~al.,}{{Garai}
  et~al.}{2016}]{Garai2}
{Garai} Z.,  et~al., 2016, \mn@doi [Astronomische Nachrichten]
  {10.1002/asna.201512310}, \href
  {https://ui.adsabs.harvard.edu/abs/2016AN....337..261G} {337, 261}

\bibitem[\protect\citeauthoryear{{Garai} et~al.,}{{Garai}
  et~al.}{2017}]{Garai1}
{Garai} Z.,  et~al., 2017, \mn@doi [Astronomische Nachrichten]
  {10.1002/asna.201613208}, \href
  {https://ui.adsabs.harvard.edu/abs/2017AN....338...35G} {338, 35}

\bibitem[\protect\citeauthoryear{{Grimm} et~al.,}{{Grimm}
  et~al.}{2018}]{Grimm1}
{Grimm} S.~L.,  et~al., 2018, \mn@doi [\aap] {10.1051/0004-6361/201732233},
  \href {https://ui.adsabs.harvard.edu/abs/2018A&A...613A..68G} {613, A68}

\bibitem[\protect\citeauthoryear{{Hartman} et~al.,}{{Hartman}
  et~al.}{2014}]{Hartman1}
{Hartman} J.~D.,  et~al., 2014, \mn@doi [\aj] {10.1088/0004-6256/147/6/128},
  \href {https://ui.adsabs.harvard.edu/abs/2014AJ....147..128H} {147, 128}

\bibitem[\protect\citeauthoryear{{Hatzes}, {Cochran}  \& {Endl}}{{Hatzes}
  et~al.}{2010}]{Hatzes1}
{Hatzes} A.~P.,  {Cochran} W.~D.,   {Endl} M.,  2010, in {Haghighipour} N.,
  ed.,  Astrophysics and Space Science Library Vol. 366, Planets in Binary Star
  Systems. p.~51, \mn@doi{10.1007/978-90-481-8687-7_3}

\bibitem[\protect\citeauthoryear{{Holman} \& {Murray}}{{Holman} \&
  {Murray}}{2005}]{Holman1}
{Holman} M.~J.,  {Murray} N.~W.,  2005, \mn@doi [Science]
  {10.1126/science.1107822}, \href
  {https://ui.adsabs.harvard.edu/abs/2005Sci...307.1288H} {307, 1288}

\bibitem[\protect\citeauthoryear{{Holman} et~al.,}{{Holman}
  et~al.}{2010}]{Holman2}
{Holman} M.~J.,  et~al., 2010, \mn@doi [Science] {10.1126/science.1195778},
  \href {https://ui.adsabs.harvard.edu/abs/2010Sci...330...51H} {330, 51}

\bibitem[\protect\citeauthoryear{{Howell} et~al.,}{{Howell}
  et~al.}{2014}]{Howell1}
{Howell} S.~B.,  et~al., 2014, \mn@doi [\pasp] {10.1086/676406}, \href
  {https://ui.adsabs.harvard.edu/abs/2014PASP..126..398H} {126, 398}

\bibitem[\protect\citeauthoryear{{Irwin}}{{Irwin}}{1952}]{Irwin1}
{Irwin} J.~B.,  1952, \mn@doi [\apj] {10.1086/145604}, \href
  {https://ui.adsabs.harvard.edu/abs/1952ApJ...116..211I} {116, 211}

\bibitem[\protect\citeauthoryear{{Knutson} et~al.,}{{Knutson}
  et~al.}{2014}]{Knutson1}
{Knutson} H.~A.,  et~al., 2014, \mn@doi [\apj] {10.1088/0004-637X/785/2/126},
  \href {https://ui.adsabs.harvard.edu/abs/2014ApJ...785..126K} {785, 126}

\bibitem[\protect\citeauthoryear{{Law} et~al.,}{{Law} et~al.}{2014}]{Law1}
{Law} N.~M.,  et~al., 2014, \mn@doi [\apj] {10.1088/0004-637X/791/1/35}, \href
  {https://ui.adsabs.harvard.edu/abs/2014ApJ...791...35L} {791, 35}

\bibitem[\protect\citeauthoryear{{Lissauer} et~al.,}{{Lissauer}
  et~al.}{2012}]{Lissauer1}
{Lissauer} J.~J.,  et~al., 2012, \mn@doi [\apj] {10.1088/0004-637X/750/2/112},
  \href {https://ui.adsabs.harvard.edu/abs/2012ApJ...750..112L} {750, 112}

\bibitem[\protect\citeauthoryear{{Maciejewski} et~al.,}{{Maciejewski}
  et~al.}{2013}]{Maciejewski1}
{Maciejewski} G.,  et~al., 2013, \mn@doi [\aap] {10.1051/0004-6361/201220739},
  \href {https://ui.adsabs.harvard.edu/abs/2013A&A...551A.108M} {551, A108}

\bibitem[\protect\citeauthoryear{{Mayor} \& {Queloz}}{{Mayor} \&
  {Queloz}}{1995}]{Mayor1}
{Mayor} M.,  {Queloz} D.,  1995, \mn@doi [\nat] {10.1038/378355a0}, \href
  {https://ui.adsabs.harvard.edu/abs/1995Natur.378..355M} {378, 355}

\bibitem[\protect\citeauthoryear{{McCullough}, {Stys}, {Valenti}, {Fleming},
  {Janes}  \& {Heasley}}{{McCullough} et~al.}{2005}]{McCullough1}
{McCullough} P.~R.,  {Stys} J.~E.,  {Valenti} J.~A.,  {Fleming} S.~W.,  {Janes}
  K.~A.,   {Heasley} J.~N.,  2005, \mn@doi [\pasp] {10.1086/432024}, \href
  {https://ui.adsabs.harvard.edu/abs/2005PASP..117..783M} {117, 783}

\bibitem[\protect\citeauthoryear{{Nascimbeni} et~al.,}{{Nascimbeni}
  et~al.}{2013}]{Nascimbeni1}
{Nascimbeni} V.,  et~al., 2013, \mn@doi [\aap] {10.1051/0004-6361/201219601},
  \href {https://ui.adsabs.harvard.edu/abs/2013A&A...549A..30N} {549, A30}

\bibitem[\protect\citeauthoryear{{Nesvorn{\'y}}, {Kipping}, {Buchhave},
  {Bakos}, {Hartman}  \& {Schmitt}}{{Nesvorn{\'y}} et~al.}{2012}]{Nesvorny1}
{Nesvorn{\'y}} D.,  {Kipping} D.~M.,  {Buchhave} L.~A.,  {Bakos} G.~{\'A}.,
  {Hartman} J.,   {Schmitt} A.~R.,  2012, \mn@doi [Science]
  {10.1126/science.1221141}, \href
  {https://ui.adsabs.harvard.edu/abs/2012Sci...336.1133N} {336, 1133}

\bibitem[\protect\citeauthoryear{{Nesvorn{\'y}}, {Kipping}, {Terrell},
  {Hartman}, {Bakos}  \& {Buchhave}}{{Nesvorn{\'y}} et~al.}{2013}]{Nesvorny2}
{Nesvorn{\'y}} D.,  {Kipping} D.,  {Terrell} D.,  {Hartman} J.,  {Bakos}
  G.~{\'A}.,   {Buchhave} L.~A.,  2013, \mn@doi [\apj]
  {10.1088/0004-637X/777/1/3}, \href
  {https://ui.adsabs.harvard.edu/abs/2013ApJ...777....3N} {777, 3}

\bibitem[\protect\citeauthoryear{{Neveu-VanMalle} et~al.,}{{Neveu-VanMalle}
  et~al.}{2016}]{Neveu-VanMalle1}
{Neveu-VanMalle} M.,  et~al., 2016, \mn@doi [\aap]
  {10.1051/0004-6361/201526965}, \href
  {https://ui.adsabs.harvard.edu/abs/2016A&A...586A..93N} {586, A93}

\bibitem[\protect\citeauthoryear{{Ofir}, {Dreizler}, {Zechmeister}  \&
  {Husser}}{{Ofir} et~al.}{2014}]{Ofir1}
{Ofir} A.,  {Dreizler} S.,  {Zechmeister} M.,   {Husser} T.-O.,  2014, \mn@doi
  [\aap] {10.1051/0004-6361/201220935}, \href
  {https://ui.adsabs.harvard.edu/abs/2014A&A...561A.103O} {561, A103}

\bibitem[\protect\citeauthoryear{{Perruchot} et~al.,}{{Perruchot}
  et~al.}{2008}]{Perruchot1}
{Perruchot} S.,  et~al., 2008, in \procspie. p. 70140J,
  \mn@doi{10.1117/12.787379}

\bibitem[\protect\citeauthoryear{{Poddan{\'y}}, {Br{\'a}t}  \&
  {Pejcha}}{{Poddan{\'y}} et~al.}{2010}]{Poddany1}
{Poddan{\'y}} S.,  {Br{\'a}t} L.,   {Pejcha} O.,  2010, \mn@doi [\na]
  {10.1016/j.newast.2009.09.001}, \href
  {https://ui.adsabs.harvard.edu/abs/2010NewA...15..297P} {15, 297}

\bibitem[\protect\citeauthoryear{{Popper} \& {Etzel}}{{Popper} \&
  {Etzel}}{1981}]{Popper1}
{Popper} D.~M.,  {Etzel} P.~B.,  1981, \mn@doi [\aj] {10.1086/112862}, \href
  {https://ui.adsabs.harvard.edu/abs/1981AJ.....86..102P} {86, 102}

\bibitem[\protect\citeauthoryear{{Pribulla}}{{Pribulla}}{2004}]{Pribulla2}
{Pribulla} T.,  2004, in {Hilditch} R.~W.,  {Hensberge} H.,   {Pavlovski} K.,
  eds,  Astronomical Society of the Pacific Conference Series Vol. 318,
  Spectroscopically and Spatially Resolving the Components of the Close Binary
  Stars. pp 117--119

\bibitem[\protect\citeauthoryear{{Pribulla}}{{Pribulla}}{2012}]{Pribulla3}
{Pribulla} T.,  2012, in {Richards} M.~T.,  {Hubeny} I.,  eds,  IAU Symposium
  Vol. 282, From Interacting Binaries to Exoplanets: Essential Modeling Tools.
  pp 279--282, \mn@doi{10.1017/S1743921311027566}

\bibitem[\protect\citeauthoryear{{Pribulla} et~al.,}{{Pribulla}
  et~al.}{2015}]{Pribulla1}
{Pribulla} T.,  et~al., 2015, \mn@doi [Astronomische Nachrichten]
  {10.1002/asna.201512202}, \href
  {https://ui.adsabs.harvard.edu/abs/2015AN....336..682P} {336, 682}

\bibitem[\protect\citeauthoryear{{Pych}}{{Pych}}{2004}]{Pych1}
{Pych} W.,  2004, \mn@doi [\pasp] {10.1086/381786}, \href
  {https://ui.adsabs.harvard.edu/abs/2004PASP..116..148P} {116, 148}

\bibitem[\protect\citeauthoryear{{Ricker} et~al.,}{{Ricker}
  et~al.}{2014}]{Ricker1}
{Ricker} G.~R.,  et~al., 2014, in \procspie. p. 914320 (\mn@eprint {arXiv}
  {1406.0151}), \mn@doi{10.1117/12.2063489}

\bibitem[\protect\citeauthoryear{{Rucinski}}{{Rucinski}}{1992}]{Rucinski1}
{Rucinski} S.~M.,  1992, \mn@doi [\aj] {10.1086/116372}, \href
  {https://ui.adsabs.harvard.edu/abs/1992AJ....104.1968R} {104, 1968}

\bibitem[\protect\citeauthoryear{{Sandford}, {Kipping}  \&
  {Collins}}{{Sandford} et~al.}{2019}]{Sandford1}
{Sandford} E.,  {Kipping} D.,   {Collins} M.,  2019, \mn@doi [\mnras]
  {10.1093/mnras/stz2350}, \href
  {https://ui.adsabs.harvard.edu/abs/2019MNRAS.489.3162S} {489, 3162}

\bibitem[\protect\citeauthoryear{{Seeliger} et~al.,}{{Seeliger}
  et~al.}{2014}]{Seeliger1}
{Seeliger} M.,  et~al., 2014, \mn@doi [\mnras] {10.1093/mnras/stu567}, \href
  {https://ui.adsabs.harvard.edu/abs/2014MNRAS.441..304S} {441, 304}

\bibitem[\protect\citeauthoryear{{Shporer} et~al.,}{{Shporer}
  et~al.}{2011}]{Shporer1}
{Shporer} A.,  et~al., 2011, \mn@doi [\aj] {10.1088/0004-6256/142/6/195}, \href
  {https://ui.adsabs.harvard.edu/abs/2011AJ....142..195S} {142, 195}

\bibitem[\protect\citeauthoryear{{Southworth}}{{Southworth}}{2015}]{Southworth2}
{Southworth} J.,  2015, {JKTLD: Limb darkening coefficients} (\mn@eprint {ascl}
  {1511.016})

\bibitem[\protect\citeauthoryear{{Southworth}, {Maxted}  \&
  {Smalley}}{{Southworth} et~al.}{2004}]{Southworth1}
{Southworth} J.,  {Maxted} P.~F.~L.,   {Smalley} B.,  2004, \mn@doi [\mnras]
  {10.1111/j.1365-2966.2004.07871.x}, \href
  {https://ui.adsabs.harvard.edu/abs/2004MNRAS.351.1277S} {351, 1277}

\bibitem[\protect\citeauthoryear{{Street} et~al.,}{{Street}
  et~al.}{2003}]{Street1}
{Street} R.~A.,  et~al., 2003, in {Deming} D.,  {Seager} S.,  eds,
  Astronomical Society of the Pacific Conference Series Vol. 294, Scientific
  Frontiers in Research on Extrasolar Planets. pp 405--408 (\mn@eprint {arXiv}
  {astro-ph/0208233})

\bibitem[\protect\citeauthoryear{{Wang} \& {Ji}}{{Wang} \& {Ji}}{2014}]{Wang2}
{Wang} S.,  {Ji} J.,  2014, \mn@doi [\apj] {10.1088/0004-637X/795/1/85}, \href
  {https://ui.adsabs.harvard.edu/abs/2014ApJ...795...85W} {795, 85}

\bibitem[\protect\citeauthoryear{{Wang} et~al.,}{{Wang} et~al.}{2018}]{Wang1}
{Wang} X.-Y.,  et~al., 2018, \mn@doi [\pasp] {10.1088/1538-3873/aab93e}, \href
  {https://ui.adsabs.harvard.edu/abs/2018PASP..130f4401W} {130, 064401}

\makeatother
\end{thebibliography}

% Alternatively you could enter them by hand, like this:
% This method is tedious and prone to error if you have lots of references
%\begin{thebibliography}{99}
%\bibitem[\protect\citeauthoryear{Author}{2012}]{Author2012}
%Author A.~N., 2013, Journal of Improbable Astronomy, 1, 1
%\bibitem[\protect\citeauthoryear{Others}{2013}]{Others2013}
%Others S., 2012, Journal of Interesting Stuff, 17, 198
%\end{thebibliography}

%%%%%%%%%%%%%%%%%%%%%%%%%%%%%%%%%%%%%%%%%%%%%%%%%%

%%%%%%%%%%%%%%%%% APPENDICES %%%%%%%%%%%%%%%%%%%%%

%\appendix

%\section{Some extra material}

%If you want to present additional material which would interrupt the flow of the main paper,
%it can be placed in an Appendix which appears after the list of references.

%%%%%%%%%%%%%%%%%%%%%%%%%%%%%%%%%%%%%%%%%%%%%%%%%%

% Don't change these lines
\bsp	% typesetting comment
\label{lastpage}
\end{document}